%% file: Lep_trans_v14.tex
\documentclass[prd,twocolumn,showpacs,amsmath,amssymb,preprintnumbers,nofootinbib,longbibliography]{revtex4-1}
\usepackage{graphicx}
\usepackage{amsmath}
\usepackage{amssymb}
\usepackage{xspace}
\usepackage{gensymb}
\usepackage{hhline}
\usepackage{multirow}
\usepackage{booktabs}
\usepackage{hyperref}
\usepackage{cleveref}
\usepackage{aas_macros}
\usepackage{soul}
\usepackage{xcolor}
\usepackage{array}
\usepackage{adjustbox}
\usepackage{cancel}

\include{defs}

\font\FermiSmallfont=cmssq8 scaled 1200

\def\LANLppthead#1{
\null 
\begin{center}\vskip -1.0truein{\hbox to 7.0truein {
\hfill
\vbox to 1in {\vfill \FermiSmallfont
              \hbox{#1}
              \vfill}
}}\vskip-0.0truein\end{center}}

\begin{document}

\preprint{LA-UR-16-29176}


\title{Lepton asymmetry, neutrino spectral distortions, and big bang
nucleosynthesis}

\author{E. Grohs$^{1}$}
\author{George M. Fuller$^2$}
\author{C. T. Kishimoto$^{2,3}$}
\author{Mark W. Paris$^4$}

\affiliation{$^{1}$Department of Physics, University of Michigan, Ann Arbor, Michigan
48109, USA}
\affiliation{$^{2}$Department of Physics, University of California,
San Diego, La Jolla, California 92093, USA}
\affiliation{$^{3}$Department of Physics and Biophysics, University of
San Diego, San Diego, California 92110, USA}
\affiliation{$^{4}$Theoretical Division, Los Alamos National
Laboratory, Los Alamos, New Mexico 87545, USA}

\date{\today}

\begin{abstract}

We calculate Boltzmann neutrino energy transport with self-consistently coupled
nuclear reactions through the weak-decoupling-nucleosynthesis epoch in an early
universe with significant lepton numbers.  We find that the presence of lepton
asymmetry enhances processes which give rise to nonthermal neutrino spectral
distortions.  Our results reveal how asymmetries in energy and entropy density
uniquely evolve for different transport processes and neutrino flavors.  The
enhanced distortions in the neutrino spectra alter the expected big bang
nucleosynthesis light element abundance yields relative to those in the
standard Fermi-Dirac neutrino distribution cases.  These yields, sensitive to
the shapes of the neutrino energy spectra, are also sensitive to the phasing of
the growth of distortions and entropy flow with time/scale factor.  We analyze
these issues and speculate on new sensitivity limits of deuterium and helium to
lepton number.

\end{abstract}



\maketitle

\section{Introduction}

In this paper we use the \burst neutrino-transport code \cite{transport_paper}
to calculate the baseline effects of out-of-equilibrium neutrino scattering on
nucleosynthesis in an early universe with a nonzero lepton number, i.e.\ an
asymmetry in the numbers of neutrinos and antineutrinos.  Our baseline
includes: a strong, electromagnetic, and weak nuclear reaction network;
modifications to the equation of state for the primeval plasma; and a Boltzmann
neutrino energy transport network.  We do not include neutrino flavor
oscillations in this work.  Our intent is to provide a coupled Boltzmann
transport and nuclear reaction calculation to which future oscillation
calculations can be compared.  In fact, the outstanding issues in achieving
ultimate precision in big bang nucleosynthesis (BBN) simulations will revolve
around oscillations and plasma physics effects.  These issues exist in both the
zero and nonzero lepton-number cases, but are more acute in the presence of an
asymmetry.

We self-consistently follow the evolution of the neutrino phase-space
occupation numbers through the weak-decoupling-nucleosynthesis epoch.  There
are many studies of the effects of lepton numbers on light element,
BBN abundance yields.  Early work
\cite{Wagoner:1966pv,1977ARNPS..27...37S} briefly explored the changes in the
helium-4 (\heiv) abundance in the presence of large neutrino degeneracies.
Later work considered how lepton numbers could influence the \heiv yield
\cite{1996PhRvD..54.2753S,1998NuPhB.534..447K} through neutrino oscillations.
In addition, other works employed lepton numbers to constrain the cosmic microwave
background (CMB) radiation energy density
\cite{2002PhRvD..65b3511H,2008JCAP...08..011S} or the sum of the light neutrino
masses \cite{2009JCAP...07..005S}.  Refs.\
\cite{2001PhRvD..64l3506K,2011JCAP...03..035M} simultaneously investigated BBN
abundances and CMB quantities using lepton numbers.  The most recent work has
used the primordial abundances to constrain lepton numbers which have been
invoked to produce sterile neutrinos through matter-enhanced
Mikheyev-Smirnow-Wolfenstein (MSW) resonances
\cite{ABFW_2005_PRD,2006PhRvD..74h5008S,2006PhRvD..74h5015C}.  Currently, our
best constraints on these lepton numbers come from comparing the
observationally-inferred primordial abundances of either \heiv or deuterium (D)
with the predicted yields of \heiv and D calculated in these models.

Previous BBN calculations with neutrino asymmetry have made the assumption that
the neutrino energy distribution functions have thermal, Fermi-Dirac (FD)
shaped forms.  In fact, we know that neutrino scattering with electrons,
positrons and other neutrinos and electron-positron annihilation produce
nonthermal distortions in these energy distributions, with concomitant effects
on BBN abundance yields \cite{transport_paper}.  Though the nucleosynthesis
changes induced with self-consistent transport are small, they nevertheless may
be important in the context of high precision cosmology.  Anticipated Stage-IV
CMB measurements \cite{cmbs4_science_book,BMW:2017} of primordial helium and
the relativistic energy density fraction at photon decoupling, coupled with the
expected high precision deuterium measurements made with future 30-meter class
telescopes
\cite{2003ApJS..149....1K,Pettini:2012yd,Cooke:2014do,2016MNRAS.455.1512C,
2016ApJ...830..148C} will provide new probes of the relic neutrino history.

In the standard cosmology with zero lepton numbers, neutrino oscillations act
to interchange the populations of electron neutrinos and antineutrinos (\nue,
\bnue) with those of muon and tau species (\num, \bnum, \nut, \bnut)
\cite{1995PhRvD..52.3184K}.  Once we posit that there are asymmetries in the
numbers of neutrinos and antineutrinos in one or more neutrino flavors, then
neutrino oscillations will largely determine the time and temperature evolution
of the neutrino energy and flavor spectra
\cite{1991ApJ...368....1S,1994PhRvD..49.2710M,1999NuPhB.538..297C,
2002PhRvD..66b5015W,2002PhRvD..66a3008A,2002NuPhB.632..363D,2010NuPhB.837...50G,
2016arXiv160801336J,2016arXiv160903200B}.  In this paper we ignore neutrino
oscillations and provide a baseline study of the relationship between neutrino
spectral distortions arising from the lengthy ($\sim10$ Hubble times) neutrino
decoupling process and primordial nucleosynthesis.  This is an extension of the
comprehensive study of this physics in the zero lepton-number case with the
\burst code \cite{transport_paper}, and in other works
\cite{1992PhRvD..46.5378D,Dolgov:1997ne,1999NuPhB.543..269D,2000NuPhB.590..539E,
2005PhRvD..71l7301S,neff:3.046,2008AIPC.1016..403S,2013PhRvD..87g3006S,
2016JCAP...07..051D}.  We will introduce alternative descriptions of the
neutrino asymmetry to study the individual processes occurring during weak
decoupling.  Our studies in this paper, together with the methods in other
works, will be important in precision calculations for gauging the effects of
flavor oscillations in the early universe.

As we develop below, a key conclusion of a comparison of neutrino-transport
effects with and without neutrino asymmetries is nonlinear enhancements of
spectral distortion effects on BBN in the former case.  This suggests that
phenomena like collective oscillations may have interesting BBN effects in full
quantum kinetic treatments of neutrino flavor evolution through the weak
decoupling epoch.

The outline of this paper is as follows.  Section \ref{sec:num_treatment} gives
the background analytical treatment of neutrino asymmetry, focusing on the
equations germane to the early universe.  Sec.\ \ref{sec:bins} presents the
rationale in picking the neutrino-occupation-number binning scheme and other
computational parameters.  We use the same binning scheme throughout this paper
as we investigate how the occupation numbers diverge from FD equilibrium,
starting in Sec.\ \ref{sec:nu_spectra}.  In Sec.\ \ref{sec:asymm}, we present a
new way of characterizing degenerate neutrinos in the early universe.  Sec.\
\ref{sec:abundances} details the changes to the primordial abundances from the
out-of-equilibrium spectra.  We give our conclusions in Sec.\ \ref{sec:concl}.
Throughout this paper we use natural units, $\hbar=c=k_B=1$, and assume
neutrinos are massless at the temperature scales of interest.

\section{Analytical Treatment}\label{sec:num_treatment}

\begin{figure}
   \begin{center}
      \includegraphics[width=\columnwidth]{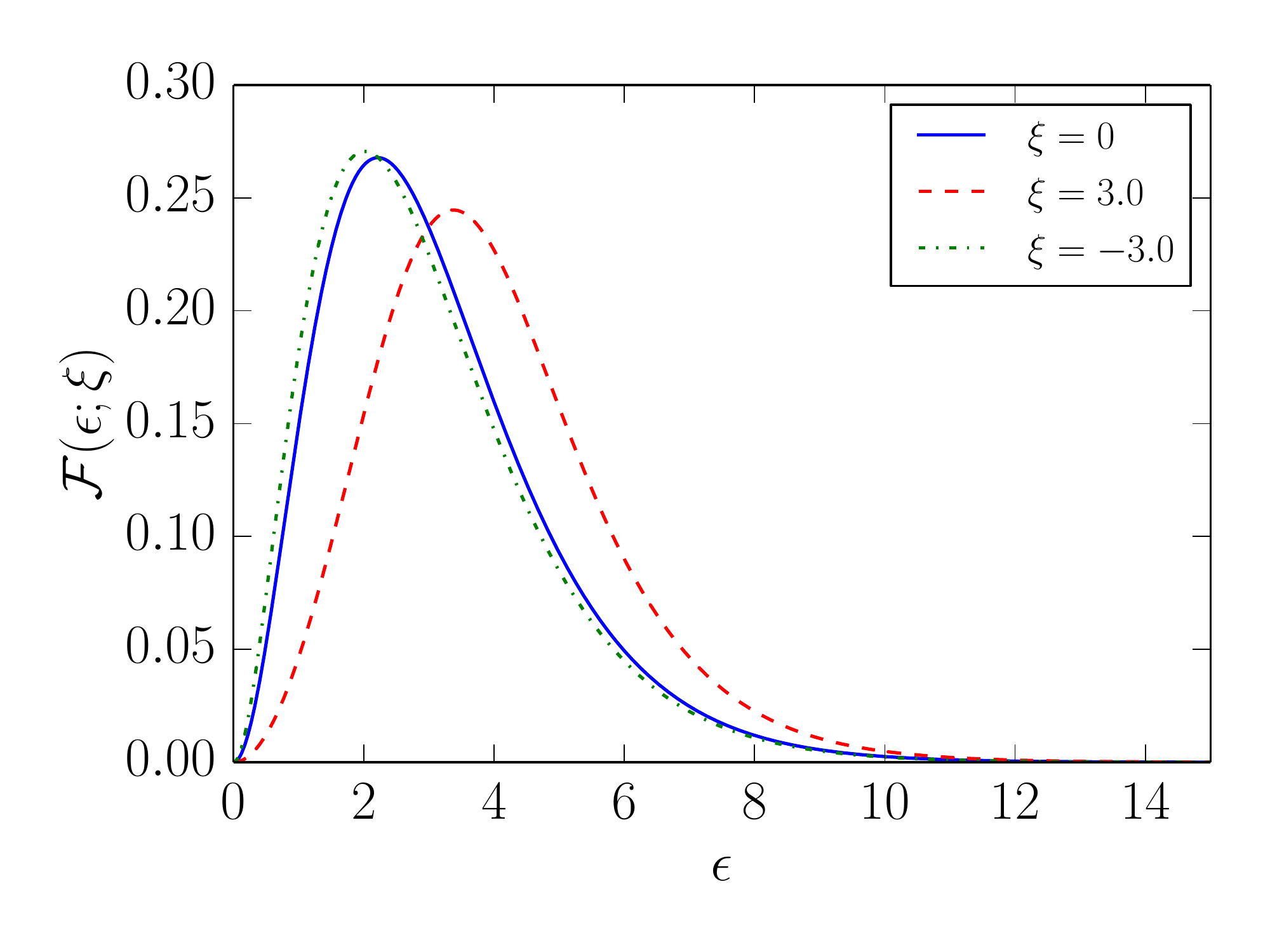}
   \end{center}
   \caption{\label{fig:deg_num} Normalized number density plotted against \eps
   for three choices of degeneracy parameter: nondegenerate ($\xi=0$, solid blue),
   degenerate with an excess ($\xi=3.0$, dashed red), and degenerate with a deficit
   ($\xi=-3.0$, dash-dotted green).
   }
\end{figure}

To characterize the lepton asymmetry residing in the neutrino seas in the early
universe, we use the following expression in terms of neutrino, $\nu$,
antineutrino, \bnu, and photon, $\gamma$, number densities to define the lepton
number for a given neutrino flavor
\begin{equation}\label{eq:lnu1}
  L_i \equiv \frac{n_{\nu_i} - n_{\bnu_i}}{n_\gamma},
\end{equation}
where $i=e,\mu,\tau$.  The photons are assumed to be in a Planck distribution
at plasma temperature $T$, with number density
\beq\label{eq:ngamma}
  n_\gamma = \frac{2\zeta(3)}{\pi^2}T^3,
\eeq
where $\zeta(3)\approx1.202$.  The neutrino spectra have general nonthermal
distributions and their number densities are given by the integration
\beq\label{eq:nnu}
  n_{\nu_i} = \frac{\tcm^3}{2\pi^2}\int_0^{\infty}d\eps\,\eps^2f_{\nu_i}(\eps).
\eeq
Here, \tcm is the comoving temperature parameter and scales inversely with
scale factor $a$
\beq
  \tcm(a)=T_{{\rm cm,}i}\left(\frac{a_i}{a}\right),
\eeq
where the $i$ subscripts reflect a choice of an initial epoch to begin the
scaling.  In this paper, we will choose $T_{{\rm cm,}i}$ such that \tcm is
coincident with the plasma temperature when $T=10\,{\rm MeV}$.  For $T>T_{{\rm
cm,}i}$, the plasma temperature and comoving temperature parameter are nearly
equal as the neutrinos are in thermal equilibrium with the
photon/electron/positron plasma.  $T$ and \tcm diverge from one another once
electrons and positrons begin annihilating into photon and
neutrino/antineutrino pairs below a temperature scale of $1\,{\rm MeV}$.  The
dummy variable \eps in Eq.\ \eqref{eq:nnu} is the comoving energy and related
to $E_\nu$, the neutrino energy, by $\epsilon=E_\nu/\tcm$.  The sets of
$f_{\nu_i}$ are the phase-space occupation numbers (also referred to as
occupation probabilities) for species $\nu_i$ indexed by \eps.  In equilibrium
the occupation numbers for neutrinos behave as FD
\beq\label{eq:feq}
  \feq(\epsilon;\xi) = \frac{1}{e^{\epsilon-\xi}+1},
\eeq
where $\xi$ is the neutrino degeneracy parameter related to the chemical
potential as $\xi=\mu/\tcm$.  Unlike the lepton number for flavor $i$ in Eq.\
\eqref{eq:lnu1}, the corresponding degeneracy parameter $\xi_i$ is a comoving
invariant.  If we consider the equilibrium occupation numbers in the expression
for number density, Eq.\ \eqref{eq:nnu}, we find
\beq
  n^{\rm (eq)}_{\nu} = \frac{\tcm^3}{2\pi^2}
  \int_0^{\infty}d\eps\frac{\eps^2}{e^{\eps-\xi}+1}
  =\frac{\tcm^3}{2\pi^2}F_2(\xi),
\eeq
where $F_2(\xi)$ is the relativistic Fermi integral given by the general expression
\beq\label{eq:F_int}
  F_k(\xi) = \int_0^{\infty}dx\frac{x^k}{e^{x-\xi}+1}.
\eeq
We can define the following normalized number distribution
\beq\label{eq:norm_nd}
  \mathcal{F}(\eps;\xi)d\eps\equiv\frac{dn}{\int dn}
  =\frac{1}{F_2(\xi)}\frac{\eps^2 d\eps}{e^{\eps-\xi}+1}.
\eeq
Figure \ref{fig:deg_num} shows $\mathcal{F}$ plotted against \eps for three
different values of $\xi$.

The expressions for the number densities in Eqs.\ \eqref{eq:ngamma} and
\eqref{eq:nnu} have different temperature/energy scales.  As the temperature
decreases, electrons and positrons will annihilate to produce photons
primarily, thereby changing $T$ with respect to \tcm.  As a result, Eq.\
\eqref{eq:lnu1} decreases from the addition of extra photons.  To alleviate
this complication, we calculate the lepton number at a high enough temperature
such that the neutrinos are in thermal and chemical equilibrium with the
plasma.  We can take $\tcm=T$ at high enough temperature and write Eq.\
\eqref{eq:lnu1} as
\begin{equation}\label{eq:lnu2}
  L^{\star}_i=\frac{1}{4\zeta(3)}\int_0^{\infty}d\epsilon\,
  \epsilon^2[f_{\nu_i}(\epsilon) - f_{\bnu_i}(\epsilon)],
\end{equation}
where we call \lstari the comoving lepton number.  Eq.\ \eqref{eq:lnu2}
simplifies further if we use the FD expression in Eq.\ \eqref{eq:feq} and
recognize that in chemical equilibrium the degeneracy parameters for neutrinos
are equal in magnitude and opposite in sign to those of antineutrinos
\beq\label{eq:lnu3}
  L^{\star}_i = \frac{\pi^3}{12\zeta(3)}\left[\frac{\xi_i}{\pi}
  + \left(\frac{\xi_i}{\pi}\right)^3\right],
\eeq
where $\xi_i$ is the degeneracy parameter for neutrinos of flavor $i$.  Eq.\
\eqref{eq:lnu3} provides an algebraic expression for relating the lepton number
to the degeneracy parameter with no explicit dependence on temperature.  We
will give our results in terms of comoving lepton number and use Eq.\
\eqref{eq:lnu3} to calculate the degeneracy parameter for input into the
computations.  In this paper, we will only consider scenarios where all three
neutrino flavors have identical comoving lepton numbers.  Unless otherwise
stated, we will drop the $i$ subscript and replace it with the neutrino symbol,
i.e.\ \lstarnu, to refer to all three flavors.

Degeneracy in the neutrino sector increases the total energy density in
radiation.  The parameter \neff is defined in terms of the plasma temperature
and the radiation energy density
\beq\label{eq:neff}
  \rho_{\rm rad}=\left[2
  +\frac{7}{4}\left(\frac{4}{11}\right)^{4/3}\neff\right]
  \frac{\pi^2}{30}T^4.
\eeq
Eq.\ \eqref{eq:neff} can be used at any epoch, even one in which there exists
seas of electrons and positrons, e.g.\ Eq.\ (31) in Ref.\
\cite{transport_paper}.  We will consider $\rho_{\rm rad}$ and $T$ at the
epoch $T=1\,{\rm keV}$, after the relic seas of positrons and electrons
annihilate.  Assuming equilibrium spectra for all neutrino species, the
deviation of \neff, $\Delta\neff$, from exactly 3 would be
\beq\label{eq:dneff1}
  \Delta\neff \equiv \neff - 3 = \sum\limits_i
  \left[\frac{30}{7}\left(\frac{\xi_i}{\pi}\right)^2
  + \frac{15}{7}\left(\frac{\xi_i}{\pi}\right)^4\right],
\eeq
where the summation assumes the possibility of different neutrino degeneracy
parameters for each flavor \cite{2000NuPhB.590..539E,Shimon:2010cb}.

We begin by presenting the case of instantaneous neutrino decoupling with pure
equilibrium FD distributions.  Table \ref{tab:lep_no_trans} shows the
deviations in energy densities for neutrinos and antineutrinos with respect to
nondegenerate FD equilibrium, the asymptotic ratio of \tcm to $T$, and the
change to \neff, for various comoving lepton numbers.  In this paper, we will
colloquially refer to the asymptote of any quantity as the ``freeze-out''
value.  For the values of \lstarnu in Table \ref{tab:lep_no_trans}, a decade
decrease in \lstarnu produces comparable decreases in $\delta\rho_\nu$ and
$|\delta\rho_{\bnu}|$.  \lstarnu is related to the energy densities through the
degeneracy parameter derived from Eq.\ \eqref{eq:lnu3}, which is approximately
linear in $\xi$ for small \lstarnu.  The change in \neff is quadratic in $\xi$
which is discernible for $\lstarnu=10^{-1}$ and $\lstarnu=10^{-2}$ at the level
of precision presented in Table \ref{tab:lep_no_trans}.  The freeze-out value
of \tcmpl is not identically $(4/11)^{1/3}$, the canonical value deduced from
covariant entropy conservation \cite{1990eaun.book.....K,2008cosm.book.....W}.
Although the neutrino-transport processes are inactive for Table
\ref{tab:lep_no_trans} and therefore the covariant entropy is conserved,
finite-temperature quantum electrodynamic (QED) effects act to perturb \tcmpl
away from the canonical value \cite{1994PhRvD..49..611H,1997PhRvD..56.5123F}. 

\begin{table*}
  \begin{center}
  \begin{tabular}{| c !{\vrule width 1.5 pt} c | c | c
  !{\vrule width 1.5 pt} c |}
    \hline
    $\lstarnu$ & $\delta\rho_{\nu}$ & $\delta\rho_{\bnu}$
    & \tcmpl & \neff\\
    \midrule[1.5pt]
    $10^{-1}$ & $0.1485$ & $-0.1300$ & $0.7149$ & $3.0479$\\ \hline
    $10^{-2}$ & $1.401\times10^{-2}$ & $-1.382\times10^{-2}$ & $0.7149$ & $3.0202$\\ \hline
    $10^{-3}$ & $1.392\times10^{-3}$ & $-1.390\times10^{-3}$ & $0.7149$ & $3.0199$\\ \hline
    $10^{-4}$ & $1.391\times10^{-4}$ & $-1.392\times10^{-4}$ & $0.7149$ & $3.0199$\\
    \midrule[1.5pt]
    $0$ & $0$ & $0$ & $0.7149$ & $3.0199$\\ \hline
  \end{tabular}
  \end{center}
  \caption{\label{tab:lep_no_trans}Observables and related quantities of
  interest for zero and nonzero comoving lepton numbers without neutrino
  transport.  Column 1 is the comoving lepton number.  Columns 2 and 3 give the
  relative changes of the $\nu$ and and \bnu energy densities compared to a FD
  energy distribution with zero degeneracy parameter at freeze-out.  Column 4
  shows the ratio of comoving temperature parameter to plasma temperature also at
  freeze-out.  For comparison, $(4/11)^{1/3}=0.7138$.  Column 5 gives \neff.
  \neff does not converge to precisely $3.0$ in the nondegenerate case due to the
  presence of finite-temperature-QED corrections to the equations of state for
  photons and electrons/positrons.
  }
\end{table*}

\section{Numerical Approach}
\label{sec:bins}

For this work, changes to the quantities of interest will be as small as a few
parts in $10^5$.  To ensure our results are not obfuscated by lack of numerical
precision, we need an error floor smaller than the numerical significance of a
given result.  In \burst, we bin the neutrino spectra in linear intervals in
\eps-space.  The binning scheme has two constraints: the maximum value of \eps
to set the range; and the number of bins over that range.  We denote the two
quantities as \epsmax and \nbins, respectively, and examine how they influence
the errors in our procedure.

The mathematical expressions for the neutrino spectra have no finite upper
limit in \eps.  We need to ensure \epsmax is large enough to encompass the
probability in the tails of the curves in Fig.\ \ref{fig:deg_num}.  As an
example, consider the normalized number density in Eq.\ \eqref{eq:norm_nd}.  We
would numerically evaluate the normalization condition as
\beq\label{eq:norm_cond}
  1 \simeq \int_0^{\epsmax}d\eps\,\mathcal{F}(\eps;\xi).
\eeq
For large $\eps$, $\mathcal{F}\sim \eps^2e^{-\eps+\xi}$, and so we exclude a
contribution to the above integral on the order of $\epsmax^2e^{-\epsmax}$ if
we take $\xi=0$.  If we are using double precision arithmetic, the contribution
becomes numerically insignificant for $\mathcal{F}(\epsmax;0)<10^{-16}$, which
corresponds to $\epsmax\simeq44$.  This value of \epsmax would seem like the
natural value to take without loss of a numerically significant contribution to
the integral in Eq.\ \eqref{eq:norm_cond}.  However, if we fix the number of
abscissa in the partition used when integrating Eq.\ \eqref{eq:norm_cond}
(i.e.\ fixing \nbins in the binned neutrino spectra), we lose precision in the
evaluation of the contribution to the integral from each bin as we increase
\epsmax.  Clearly, there is a trade off between \epsmax and \nbins.

\begin{figure}
   \begin{center}
      \includegraphics[width=\columnwidth]{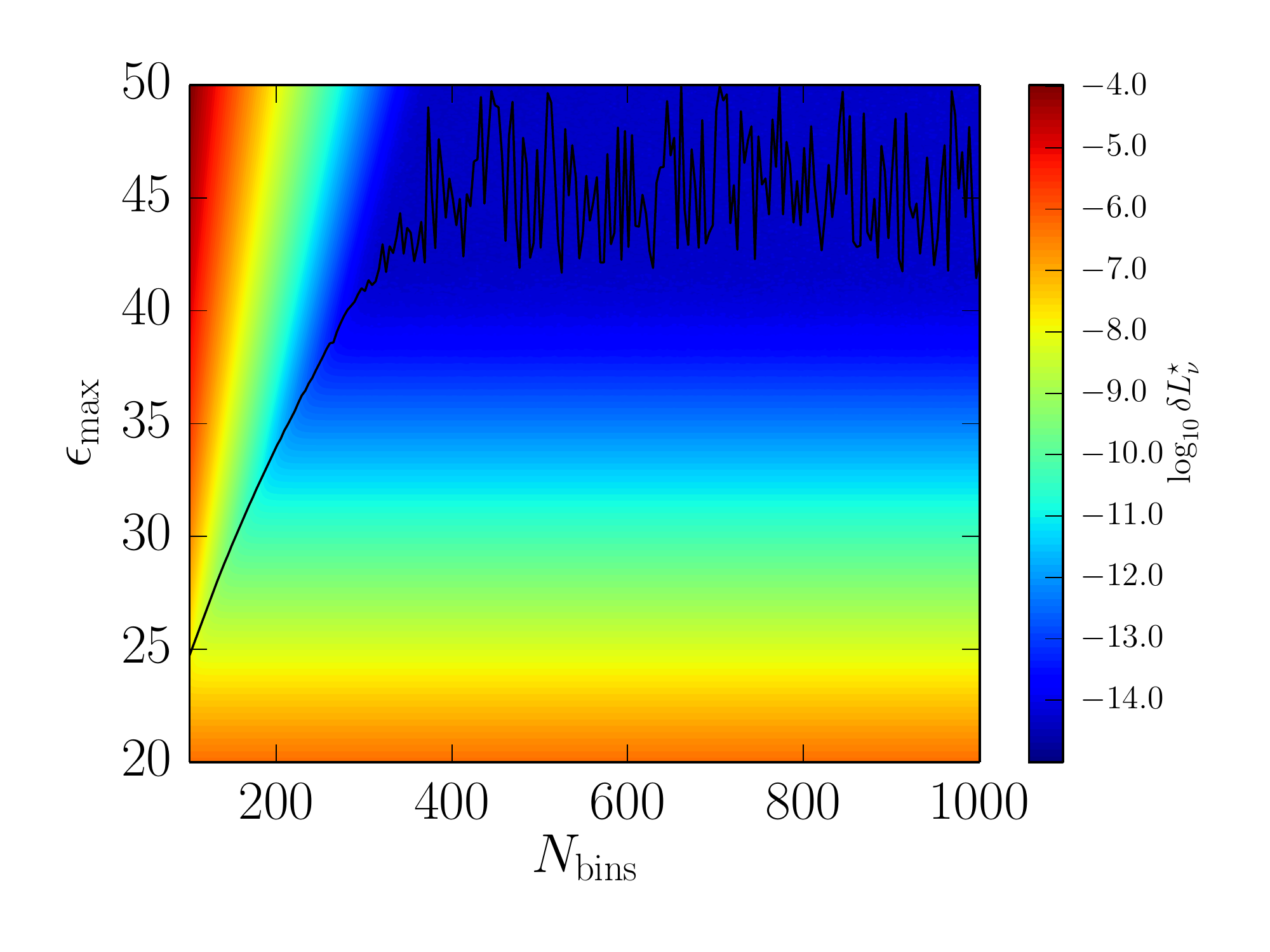}
   \end{center}
   \caption{\label{fig:contour} \epsmax versus \nbins for contours of constant
   error in \lstarnu.  The exact value of \lstarnu is $0.1$.  The black line on
   the contour space gives the value of \epsmax with the smallest error as a
   function of \nbins.
   }
\end{figure}

Figure \ref{fig:contour} examines the \epsmax versus \nbins parameter space by
looking at the calculation of the equilibrium comoving lepton number, in a
scenario where $\xi\ne0$.  We take \lstarnu to be exactly $0.1$ and solve the
cubic equation in Eq.\ \eqref{eq:lnu3} for $\xi$.  Next, we calculate neutrino
and antineutrino spectra with the equal and opposite degeneracy parameters.  We
proceed to integrate Eq.\ \eqref{eq:lnu2} with the two spectra for different
pairs of $(\nbins,\epsmax)$ values.  The integration is carried out using
Boole's rule, a fifth-order integration method for linearly spaced abscissas.
Figure \ref{fig:contour} shows the filled contours of $\log_{10}$ values for
the error in \lstarnu
\beq
  \delta\lstarnu = \biggr|\frac{\lstarnu[\epsmax,\nbins] - 0.1}{0.1}\biggr|,
\eeq
for a given pair $(\nbins,\epsmax)$.  We immediately see the loss of precision
in the upper-left corner of the parameter space, corresponding to small \nbins
and large \epsmax.  Furthermore, for $\epsmax\lesssim40$, the error value
flat lines with increasing \nbins, implying that the error is a result of a
too small choice for \epsmax.  The black curve superimposed on the heat map
gives the value of \epsmax with the lowest error as a function of \nbins.  It
monotonically increases for $100<\nbins\lesssim300$, at which point it reaches
$\epsmax\sim40$ and begins to fluctuate.  The fluctuations are a result of
reaching the double-precision floor, implying that increasing \nbins adds no
more numerical significance.

The computation time required to run \burst scales as $\nbins^3$.  In this
paper, we attempt to be as comprehensive as possible when exploring neutrino
energy transport with nonzero lepton numbers.  Therefore, we will choose $100$
bins for the sake of expediency.  Fig.\ \ref{fig:contour} guides us in picking
$\epsmax=25$, and dictates a floor of $\sim10^{-8}$ for our best possible
precision.  It would appear that the choice $(\nbins,\epsmax)\simeq(300,40)$
would give the absolute best precision for calculating \lstarnu.  This is valid
if using the linearly spaced abscissas as a binning scheme.  We highlight both
the precision and timing needs for a comprehensive numerical study on binning
schemes.  Such a study would be germane for the more general problem which
includes neutrino oscillations and disparate lepton numbers in the three active
species \cite{2016JCAP...07..051D,2016arXiv160903200B,2016arXiv160801336J}.

For more details on the numerics of \burst, we refer the reader to Ref.\
\cite{transport_paper}.  We have essentially preserved the computational
parameters except for a quantity related to the determination of nonzero
scattering rates.  Ref.\ \cite{transport_paper} used $\nuratiotol=30$, and in
this work, we use $\nuratiotol=3$.

\section{Neutrino Spectra}
\label{sec:nu_spectra}

In this section we give a detailed accounting of how the neutrino energy
spectra evolve through weak decoupling in the presence of zero and nonzero
lepton numbers.  In the first subsection we integrate the complete transport
network, including all the neutrino scattering processes in Table I of Ref.\
\cite{transport_paper}, from a comoving temperature parameter $\tcm=10\,{\rm
MeV}$ down to $\tcm=15\,{\rm keV}$.  In the second subsection we investigate
how the different interactions between neutrinos and charged leptons affect the
spectra.

We compare our results to that of FD equilibrium.  For the neutrino occupation
numbers, we use the following notation to characterize the deviations from FD
equilibrium
\beq\label{eq:delta_f_xi}
  \delta f_\xi(\epsilon) = \frac{f(\epsilon) - \feq(\epsilon;\xi)}{\feq(\epsilon;\xi)}.
\eeq
Here, $\feq(\epsilon;\xi)$ is the FD equilibrium occupation number for degeneracy
parameter $\xi$ given in Eq.\ \eqref{eq:feq}.  When it is obvious, we will drop
the argument $\epsilon$, i.e.\ $\delta f_\xi(\epsilon)\rightarrow\delta f_\xi$.
As an example, $\delta f_0$ gives the relative difference of the occupation
number from the nondegenerate, zero chemical potential FD equilibrium value.

We also examine the absolute changes for the number and energy distributions
\begin{align}
  \Delta\left(\frac{dn}{d\epsilon}\right)_\xi\equiv\frac{\tcm^3}{2\pi^2}
  \epsilon^2[f(\epsilon) - \feq(\epsilon;\xi)] &&\quad {\rm (number)},\\
  \Delta\left(\frac{d\rho}{d\epsilon}\right)_\xi\equiv\frac{\tcm^4}{2\pi^2}
  \epsilon^3[f(\epsilon) - \feq(\epsilon;\xi)] &&\quad {\rm (energy)}.
\end{align}
When using the absolute change expressions, we normalize with respect to an
equilibrium number or energy density in order to compare to dimensionless
expressions.  For the energy density, we use the appropriate degeneracy factor
\beq
  \rho_\xi\equiv\frac{\tcm^4}{2\pi^2}\int_0^{\infty}d\epsilon\,
  \epsilon^3\feq(\epsilon;\xi).
\eeq
For the number density, we will exclusively use zero for the degeneracy factor
\begin{align}
  n_0&\equiv\frac{\tcm^3}{2\pi^2}\int_0^{\infty}d\epsilon\,\epsilon^3\feq(\epsilon;0)\nonumber\\
  &=\frac{3}{4}\frac{\zeta(3)}{\pi^2}\tcm^3.\label{eq:n0}
\end{align}
The out-of-equilibrium evolution of the neutrino occupation numbers driven by
scattering and annihilation processes with charged leptons does not proceed in
a unitary fashion.  Consequently, the total comoving neutrino number density
increases.  The increase in number results in an increase in energy density,
and so we use $\rho_\xi$ to normalize the absolute changes in differential
energy density distribution to compare with the initial distribution at high
temperature.  However, the difference in number density between neutrinos and
antineutrinos, characterized by the comoving lepton number in Eq.\
\eqref{eq:lnu2}, does not change with kinematic neutrino transport.  In
practice, \burst follows the evolution of neutrino and antineutrino occupation
numbers separately, precipitating the possibility of numerical error.  We will
use the same normalization for neutrino and antineutrino differential number
density distributions to study the relative error in $L^{\star}_i$.  We will
take the normalization quantity to be that of the nondegenerate number density
in Eq.\ \eqref{eq:n0}.

\subsection{All processes}

Table \ref{tab:lep_w_trans} shows how neutrino transport alters neutrino energy
densities, \neff, the ratio of comoving temperature parameter to plasma
temperature, and entropy per baryon in the plasma, \spl.  These quantities are
computed for a range of \lstari values and refer to the results at the end of
the transport calculation, $\tcm\sim1\,{\rm keV}$, well after weak decoupling.
In this table, we focus on the energy-derived quantities.  The relative changes
in energy density are with respect to a nondegenerate FD distribution at the
same comoving temperature, i.e.\
\begin{equation}
  \delta\rho_i = \frac{\displaystyle\frac{\tcm^4}{2\pi^2}\int_0^{\infty}d\epsilon\,
  \epsilon^3f_i(\epsilon) - \rho_0}{\rho_0},
  \quad \rho_0=\frac{7}{8}\frac{\pi^2}{30}\tcm^4.
\end{equation}
Columns 2 - 5 of Table \ref{tab:lep_w_trans} show the relative changes in
energy density at $\tcm=1\,{\rm keV}$, once the neutrino spectra have converged
to their out-of-equilibrium shapes.  We see a monotonic decrease in
$\delta\rho_\nu$ for the neutrinos, and a monotonic increase in
$\delta\rho_{\bnu}$ for the antineutrinos with decreasing \lstarnu.  Column 6
gives the ratio of \tcmpl at the end of the simulation.  \tcmpl increases with
decreasing lepton number.  However, the decrease is less than one part in
$10^5$ between $\lstarnu=0.1$ and $\lstarnu=0$.  The larger lepton number
implies a larger total energy density which increases the Hubble expansion
rate.  The faster expansion implies a smaller time window for the entropy flow
out of the plasma and into the neutrino seas.  As a result, the evolution of
the plasma temperature is such that larger lepton numbers will maintain $T$ at
higher values, and the ratio \tcmpl at freeze-out will decrease, albeit by an
amount which is numerically insignificant.  With the changes in energy
densities and temperature ratios, we can calculate \neff
\beq\label{eq:neff_pert}
  \neff = \left[\frac{\tcmpl}{(4/11)^{1/3}}\right]^4
  \frac{1}{2}[(2+\delta\rho_{\nu_e}+\delta\rho_{\bnue})
  + 2(2+\delta\rho_{\num}+\delta\rho_{\bnum})].
\eeq
The coefficient in front of the second parenthetical expression, equal in value
to $2$, results from the approximation in taking the $\mu$ and $\tau$ flavors
to behave identically.  The approximation employed here is valid as there are
no $\mu$ and $\tau$ charged leptons in the plasma and $L_i$ is the same in all
flavors.
Both Refs.\ \cite{neff:3.046,2016JCAP...07..051D} calculate weak decoupling
with a network featuring neutrino flavor oscillations, which are absent in our
calculation in Table \ref{tab:lep_w_trans}.  However, Ref.\
\cite{2016JCAP...07..051D} states that oscillations have no affect on the value
of \neff at the level of precision which they use.  The difference in our value
of \neff versus the standard calculation of Ref.\ \cite{neff:3.046} is most
likely due to a different implementation of the finite-temperature QED effects
detailed in Refs.\ \cite{1994PhRvD..49..611H,1997PhRvD..56.5123F}.  Ref.\
\cite{neff:3.046} uses the perturbative approach outlined in Ref.\
\cite{Mangano:3.040} compared to our nonperturbative approach.  We leave a
detailed study of the finite-temperature-QED-effect numerics to future work.

The final column of Table \ref{tab:lep_w_trans} shows the change in the entropy
per baryon in the plasma.  The relative changes in entropy for varying lepton
numbers are large enough to see a difference at the level of precision Table
\ref{tab:lep_w_trans} uses, unlike \tcmpl.  With the faster expansion,
neutrinos have less time to interact with the plasma, yielding a smaller
entropy flow.

\begin{table*}
  \begin{center}
  \begin{tabular}{| c !{\vrule width 1.5 pt} c | c | c | c | c
  !{\vrule width 1.5 pt} c | c |}
    \hline
    \lstarnu & $\delta\rho_{\nue}$
    & $\delta\rho_{\bnue}$
    & $\delta\rho_{\num}$
    & $\delta\rho_{\bnum}$ & \tcmpl & \neff
    & $10^3\times(\sisf - 1)$\\ 
    \midrule[1.5pt]
    $10^{-1}$ & $0.1576$ & $-0.1213$ & $0.1522$ & $-0.1265$ &
    $0.7159$ & $3.0800$ & $3.809$\\ \hline
    $10^{-2}$ & $2.298\times10^{-2}$ & $-4.888\times10^{-3}$ &
    $1.760\times10^{-2}$ & $-1.025\times10^{-2}$ & $0.7159$ &
    $3.0519$ & $3.814$\\ \hline
    $10^{-3}$ & $1.035\times10^{-2}$ & $7.564\times10^{-3}$ &
    $4.979\times10^{-3}$ & $2.194\times10^{-3}$ & $0.7159$ &
    $3.0516$ & $3.815$\\ \hline
    $10^{-4}$ & $9.096\times10^{-3}$ & $8.817\times10^{-3}$ &
    $3.725\times10^{-3}$ & $3.446\times10^{-3}$ & $0.7159$ &
    $3.0516$ & $3.815$\\ 
    \midrule[1.5pt]
    $0$ & $8.957\times10^{-3}$ & $8.957\times10^{-3}$ &
    $3.585\times10^{-3}$ & $3.585\times10^{-3}$ & $0.7159$ &
    $3.0516$ & $3.815$\\ \hline
  \end{tabular}
  \end{center}
  \caption{\label{tab:lep_w_trans} Observables and related quantities of
  interest in zero and nonzero lepton-number scenarios with neutrino transport.
  Column 1 is the lepton number.  Columns 2, 3, 4, and 5 give the relative
  changes of the \nue, \bnue, \num and \bnum energy densities compared to a FD
  energy distribution with zero degeneracy parameter.  Comparisons are given for
  $\tcm\sim 1\,{\rm keV}$, after the conclusion of weak decoupling.  Column 6
  shows the ratio of comoving temperature parameter to plasma temperature also at
  the conclusion of weak decoupling.  For comparison, $(4/11)^{1/3}=0.7138$.
  Column 7 gives \neff as calculated by Eq.\ \eqref{eq:neff_pert}.  Column 8
  gives the fractional change in the entropy per baryon in the plasma, \spl.
  }
\end{table*}

\begin{figure}
   \begin{center}
      \includegraphics[width=\columnwidth]{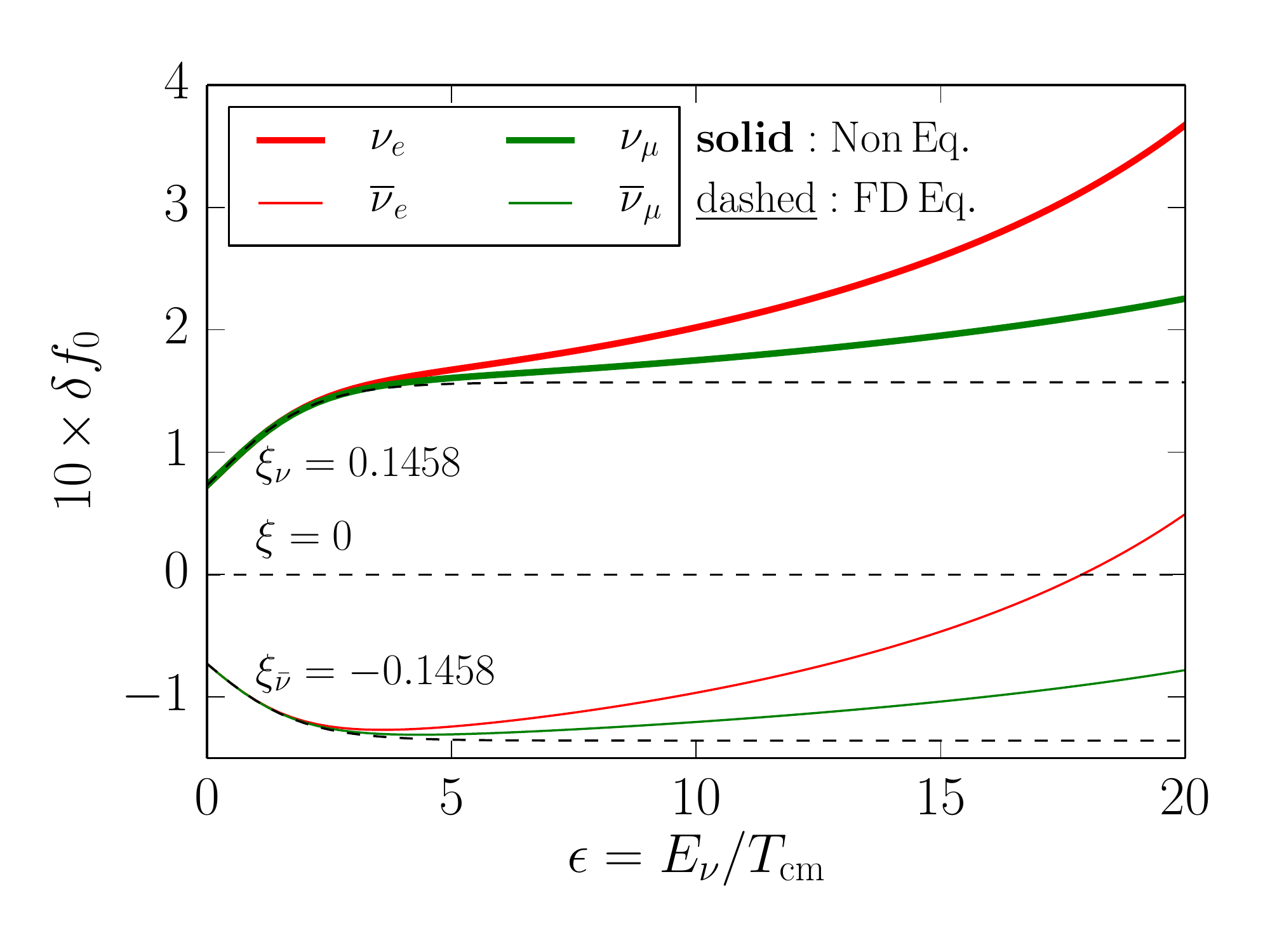}
   \end{center}
   \caption{\label{fig:occ_0_eps}Relative differences in neutrino/antineutrino
   occupation numbers plotted against \eps at $\tcm=1\,{\rm keV}$.  The relative
   differences are with respect to FD with zero degeneracy parameter.  The solid lines show
   the evolution for a scenario where $\lstarnu=0.1$.  \nue and \bnue curves are
   colored red, and \num and \bnum curves are colored green.  Neutrinos have thick
   line widths and antineutrinos have thin line widths.  Plotted for comparison are
   black dashed curves representing the equilibrium relative differences.  The top
   dashed curve corresponds to a $\nu$ spectrum with $\xi_\nu=0.1458$, the middle
   horizontal curve corresponds to a $\nu$ spectrum with $\xi_\nu=0$, and the bottom
   dashed curve corresponds to a \bnu spectrum with $\xi_{\bnu}=-0.1458$.
   }
\end{figure}

An increase in lepton number implies a larger energy density for the neutrinos
over the antineutrinos.  Fig.\ \ref{fig:occ_0_eps} shows four neutrino spectra
after the conclusion of weak decoupling in a scenario where $\lstarnu=0.1$.
Plotted against \eps is the relative difference in the neutrino occupation
number with respect to a nondegenerate spectrum.  As seen in the first data row
of Table \ref{tab:lep_w_trans}, $\delta\rho_{\nue}$ obtains the largest
difference from equilibrium.  The thick red line in Fig.\ \ref{fig:occ_0_eps}
shows the final out-of-equilibrium spectrum for \nue.  The \nue spectrum has
the largest deviation from equilibrium, congruent with Table
\ref{tab:lep_w_trans}.  The black dashed lines show equilibrium spectra for
nondegenerate (flat, horizontal line) and degenerate cases.  As \eps increases,
the neutrino curves diverge from the positive $\xi_\nu$ spectrum in much the
same manner as the antineutrino curves diverge from the negative $\xi_{\bnu}$.
The primary difference in the out-of-equilibrium spectra is due to the initial
condition that the neutrinos have larger occupation numbers over the
antineutrinos for a positive lepton number.

\begin{figure}
   \begin{center}
      \includegraphics[width=\columnwidth]{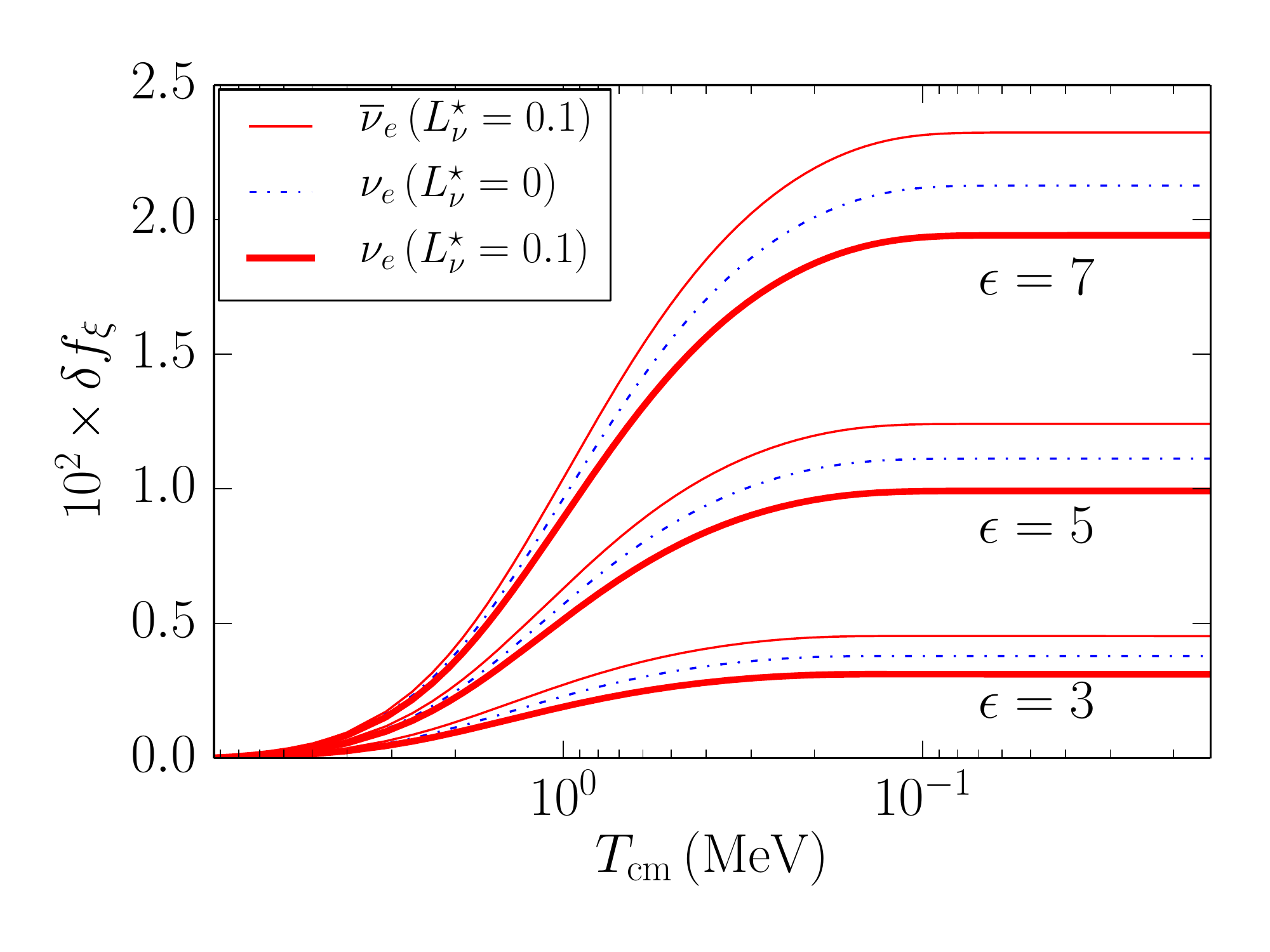}
   \end{center}
   \caption{\label{fig:occ_xi_tcm}Relative differences in electron
   neutrino/antineutrino occupation numbers plotted against \tcm.
   The relative differences are with respect to FD with the same degeneracy
   parameters as Fig.\ \ref{fig:occ_0_eps}.
   The solid lines show the evolution for a
   scenario where $\lstarnu=0.1$.  The \bnue (thin red curves) has a larger relative
   change than the \nue (thick red curves).  Plotted for comparison is the
   relative difference for \nue in a $\lstarnu=0$ scenario (blue dash-dot curves).
   The relative differences are plotted for three values of $\epsilon$,
   from bottom to top: $\epsilon=3,\,5,\,7$.
   }
\end{figure}

\begin{figure}
   \begin{center}
      \includegraphics[width=\columnwidth]{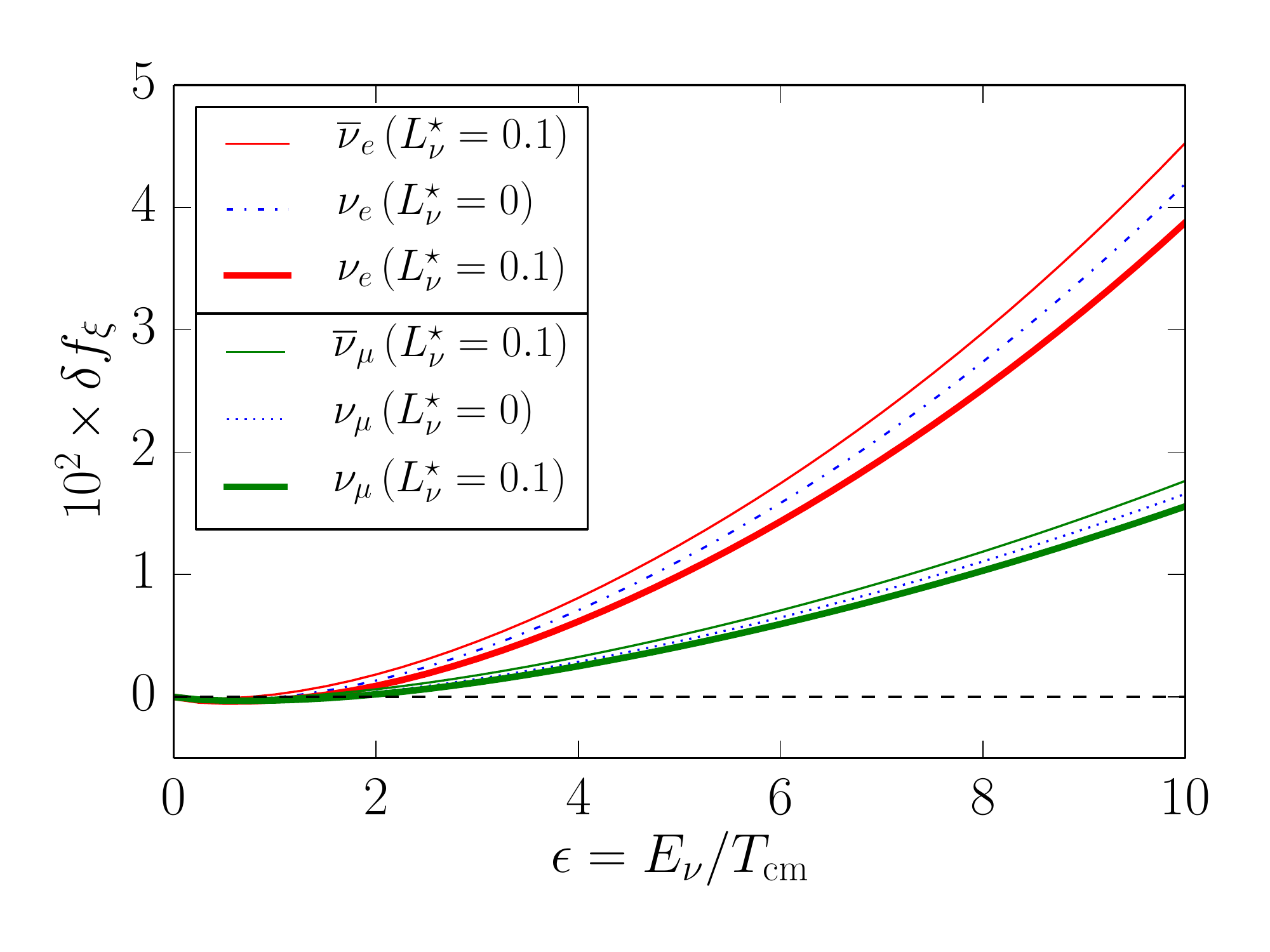}
   \end{center}
   \caption{\label{fig:occ_xi_eps} Relative differences in
   neutrino/antineutrino occupation numbers plotted against \eps at $\tcm=1\,{\rm
   keV}$.  The relative differences are with respect to FD with the same corresponding
   degeneracy parameters as Fig.\ \ref{fig:occ_0_eps}.  The solid lines are
   for a scenario where $\lstarnu=0.1$.  Plotted for comparison is the relative
   difference for \nue (blue dash-dot curve) and for \num (blue dotted curve) in a
   $\lstarnu=0$ scenario.
   }
\end{figure}

We would like to compare the out-of-equilibrium spectra to their respective
equilibrium spectra.  Such a comparison allows us to examine how the initial
asymmetry propagates through the Boltzmann network.  Fig.\ \ref{fig:occ_xi_tcm}
shows the \tcm evolution of $\delta f_\xi$ for \nue (thick solid lines) and
\bnue (thin solid lines) for a scenario where $\lstarnu=0.1$.  We only show the
relative differences for three unique values of \eps, namely
$\epsilon=3,\,5,\,7$.  The \num and \bnum spectra follow similar shapes, but
are suppressed relative to the electron flavors.  For comparison, we also plot
the out-of-equilibrium spectrum for \nue in the case of no initial asymmetry,
i.e.\ \lstarnu identically zero.
It is unnecessary to show the spectrum for \bnue when $\lstarnu=0$ because it
is exceedingly near the \nue spectrum (see Fig.\ [3] of Ref.\
\cite{transport_paper}).  For the degenerate spectra, the \bnue show a larger
divergence from equilibrium than the \nue at these three specific \epsvals.
This is consistent with Ref.\ \cite{2000NuPhB.590..539E} (see Figs.\ 8 and 9
therein) and is the case for all \eps after the neutrino spectra have frozen
out. Fig.\ \ref{fig:occ_xi_eps} shows the final freeze-out values of the
relative changes in the neutrino occupation numbers as a function of \eps.
Fig.\ \ref{fig:occ_xi_eps} is similar to Fig.\ \ref{fig:occ_0_eps} except for
the use of the general $\delta f_\xi$ instead of $\delta f_0$.  We have also
included the transport-induced out-of-equilibrium spectra for \nue and \num in
the nondegenerate scenario.  For a given flavor, the relative changes in the
nondegenerate spectrum are nearly averages of those in the $\nu$ and $\bnu$
spectra.  We also note that for $\epsilon\lesssim2$, all of the relative
differences are negative, although this is obscured in Fig.\
\ref{fig:occ_xi_eps} due to the clustering of lines.  For small \eps, the
antineutrino occupation numbers are larger than those of the neutrino, i.e.\
the $\delta f_\xi^{(\bnu)}$ are not as negative.

\begin{figure}
   \begin{center}
      \includegraphics[width=\columnwidth]{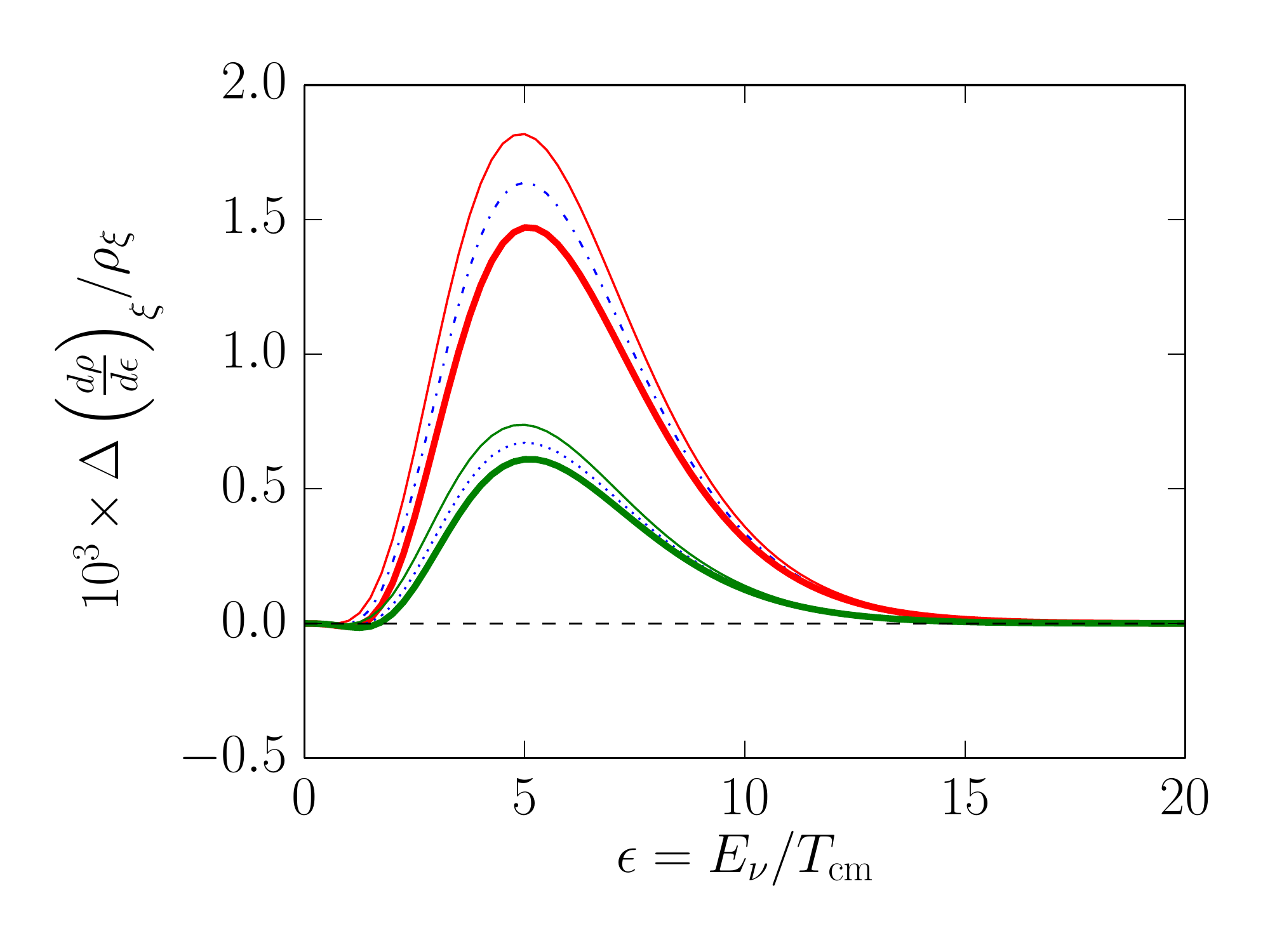}
   \end{center}
   \caption{\label{fig:rho_xi_eps}Absolute change in the neutrino/antineutrino
   energy distributions plotted against \eps at $\tcm=1\,{\rm keV}$.  The changes
   are with respect to the same degeneracy parameters as those in Fig.\
   \ref{fig:occ_xi_eps}.  Furthermore, the line colors and styles correspond to
   the same species and scenarios as Fig.\ \ref{fig:occ_xi_eps}.
   }   
\end{figure}

The weak interaction cross sections scale as $\sigma\sim G_F^2E^2$, where $G_F$
is the Fermi constant ($G_F\approx1.166\times10^{-11}\,{\rm MeV}^{-2}$) and $E$
is the total lepton energy.  We would expect a larger difference from
equilibrium for increasing \eps.  Except for the range $0<\epsilon\lesssim1$,
Figures \ref{fig:occ_xi_tcm} and \ref{fig:occ_xi_eps} clearly show an increase.
The change in the energy distribution does not follow from a scaling relation.
Fig.\ \ref{fig:rho_xi_eps} shows the normalized absolute difference in the
energy distribution plotted against \eps at the conclusion of weak decoupling.
The nomenclature for the six lines in Fig.\ \ref{fig:rho_xi_eps} is identical
to that of Fig.\ \ref{fig:occ_xi_eps}.  The energy distributions all show a
maximum at $\epsilon\sim5$.  Similar to Fig.\ \ref{fig:occ_xi_eps}, the
nondegenerate curves of Fig.\ \ref{fig:rho_xi_eps} appear to be averages of the
$\nu$ and \bnu curves in the degenerate scenario.

\begin{figure}
   \begin{center}
      \includegraphics[width=\columnwidth]{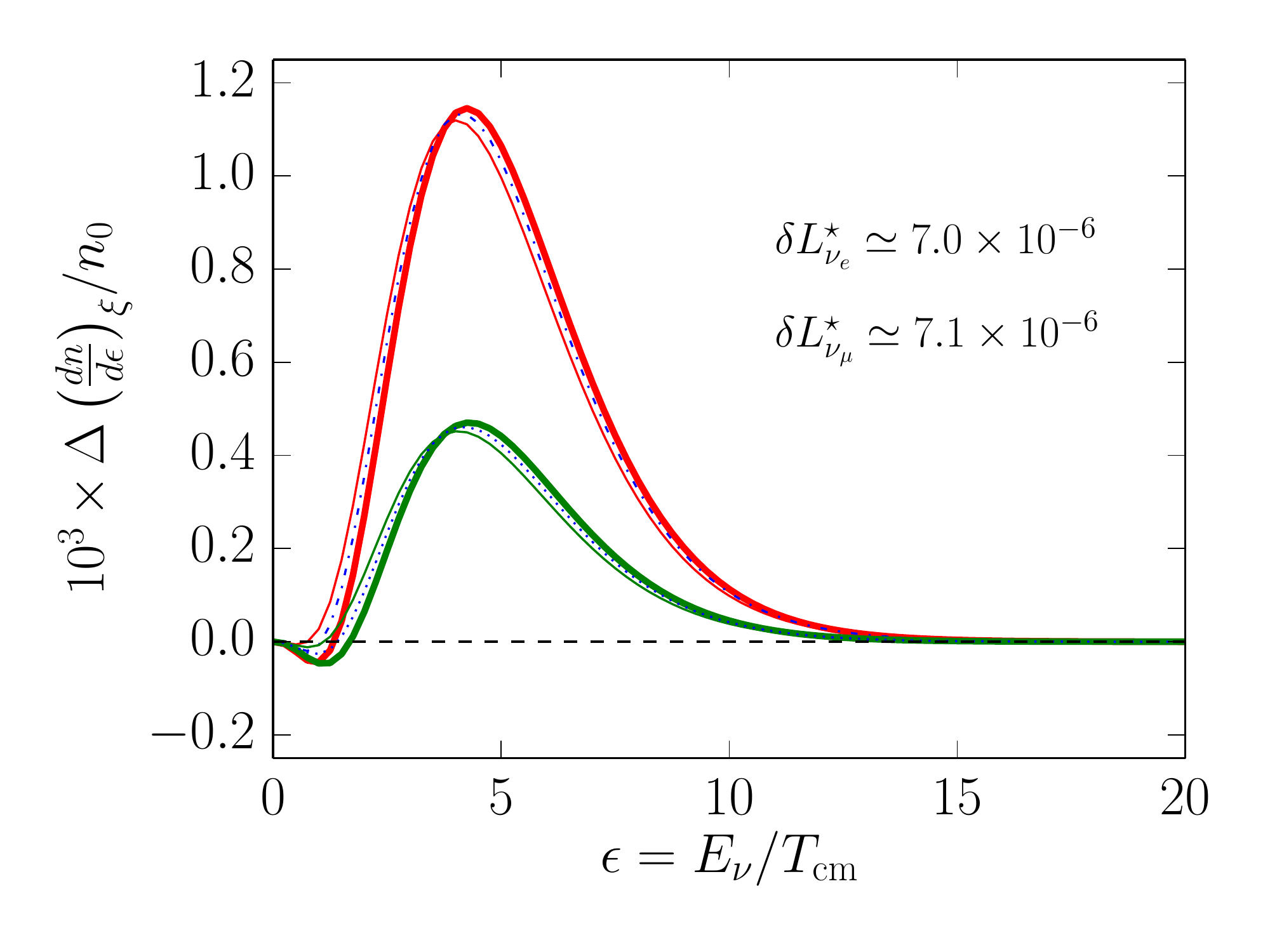}
   \end{center}
   \caption{\label{fig:num_0_eps}Absolute change in the neutrino/antineutrino
   number distributions plotted against \eps at $\tcm=1.0\,{\rm keV}$.  The
   changes are with respect to the same degeneracy parameters, and the
   nomenclature of line colors and line styles is the same as those in Fig.\
   \ref{fig:occ_xi_eps}.  The numbers given on the plot ($\delta \lstari$ for
   $i=e,\mu$) show the relative error accumulated over the course of a simulation.
   }
\end{figure}

In the positive lepton-number scenarios, the \bnu always have larger occupation numbers
than the $\nu$, when compared against the equilibrium degenerate
spectrum/distribution.  This is not surprising as the occupation numbers for
antineutrinos are suppressed, implying less blocking.  When compared against its
equilibrium distribution, the \bnu have larger rates, leading to a larger distortion.
In Fig.\ \ref{fig:num_0_eps}, we compare the out-of-equilibrium
number density distributions with those of the nondegenerate case solely.  In
other words, the normalizing factor $n_0$ is the same for each of the six
curves in Fig.\ \ref{fig:num_0_eps}.  We have adopted this nomenclature for the
comparison of number density distributions to study the change in the comoving lepton
number. None of the weak decoupling processes modify the lepton number in our
model.  The total change in number density for $\nu$ should be identical to the
total change in number density for $\overline{\nu}$.  Fig.\ \ref{fig:num_0_eps}
shows this indirectly.  We can see a difference; the $\nu$ curves are
skewed to higher \eps and have a larger maximum than the \bnu.  The negative change in the
distributions for the range $0\le\eps\lesssim2$ is much more noticeable in
Fig.\ \ref{fig:num_0_eps} than in Fig.\ \ref{fig:occ_xi_eps}.  It is clear that
the changes in \bnu become positive for smaller \eps than those of $\nu$,
implying there are more \bnu than $\nu$ for $\epsilon\lesssim2$.  Overall, when
integrating the curves in Fig.\ \ref{fig:num_0_eps}, the total changes in
number density for $\bnu$ should be the same as for $\nu$.  We have calculated
this quantity and expressed it as a relative change in the \lstari, taken
to be exactly $0.1$
\beq\label{eq:delta_l}
  \delta \lstari \equiv\frac{\displaystyle\frac{1}{4\zeta(3)}\int_0^{\infty}
  d\epsilon\,\epsilon^2[f_{\nu_i}(\eps) - f_{\bnu_i}(\eps)] - 0.1}{0.1}.
\eeq
Eq.\ \eqref{eq:delta_l} gives the relative error in our calculation.  We
conserve the comoving lepton number for both electron and muon flavor at approximately
$7\times10^{-6}$.  Also plotted in Fig.\ \ref{fig:num_0_eps} are the absolute
changes for \nue and \num in the nondegenerate scenario.  We do not directly
compare the lepton-number relative errors as the quantity is not defined for
the symmetric case.  We do note that the nondegenerate curves are close to the
average of the $\nu$ and $\bnu$ distributions, similar to that of Figs.\
\ref{fig:occ_xi_eps} and \ref{fig:rho_xi_eps}.

\begin{figure}
   \begin{center}
      \includegraphics[width=\columnwidth]{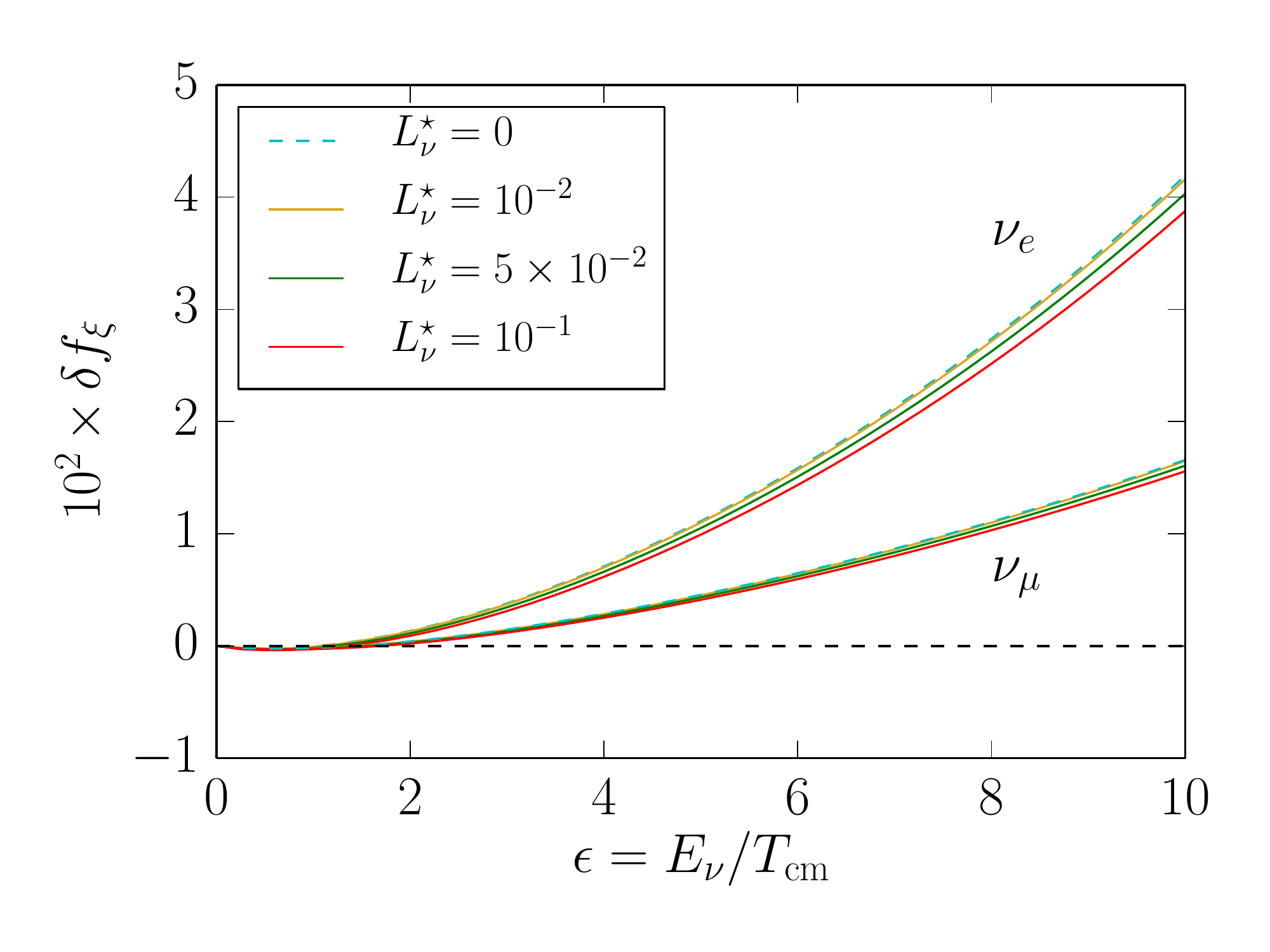}
   \end{center}
   \caption{\label{fig:occ_xi_lepnums} Relative differences in neutrino
   occupation numbers plotted against \eps at $\tcm=1\,{\rm keV}$ for various
   values of \lstarnu.  The relative differences are with respect to FD with the corresponding
   degeneracy parameter, namely $\xi=0.1458$ ($\lstarnu=10^{-1}$), $\xi=0.0730$
   ($\lstarnu=5\times10^{-2}$), $\xi=0.0146$ ($\lstarnu=10^{-2}$), and $\xi=0$ ($\lstarnu=0$).
   Shown are two sets of curves: the set with the larger relative differences
   correspond to the \nue spectral distortions, and the set with the smaller
   differences are \num.  Within each set, the increase in \lstarnu leads to a
   decrease in $\delta f_\xi$.  For the antineutrinos, the relative differences
   behave in the opposite manner: increase in \lstarnu leads to an increase in $\delta
   f_\xi$.
   }
\end{figure}

In Figs.\ \ref{fig:occ_0_eps} through \ref{fig:num_0_eps}, we have only
presented the $\lstarnu=0.1$ scenario.  Fig.\ \ref{fig:occ_xi_lepnums} shows
the relative differences in occupation number for $\nu$ plotted against \eps
for other values of \lstarnu.  The behavior of each curve is in line with those
of Fig.\ \ref{fig:occ_xi_eps}.  Not plotted are the curves for \bnu.  They also
behave in a similar manner, where $\delta f_{\bnu}$ becomes larger than $\delta
f_0$ for increasing \eps.  The result is that with transport, \lstarnu acts to
{\it increase} the asymmetries in the occupation numbers, which manifest in
differences in the absolute changes of the differential energy density.

\subsection{Individual processes}

Figs.\ \ref{fig:occ_xi_eps} and \ref{fig:rho_xi_eps} demonstrate that the
initial asymmetry in the neutrino energy density is maintained and even amplified by
scattering processes.  We can dissect the relative contribution of various
scattering processes to this amplification.

\begin{figure}
   \begin{center}
      \includegraphics[width=\columnwidth]{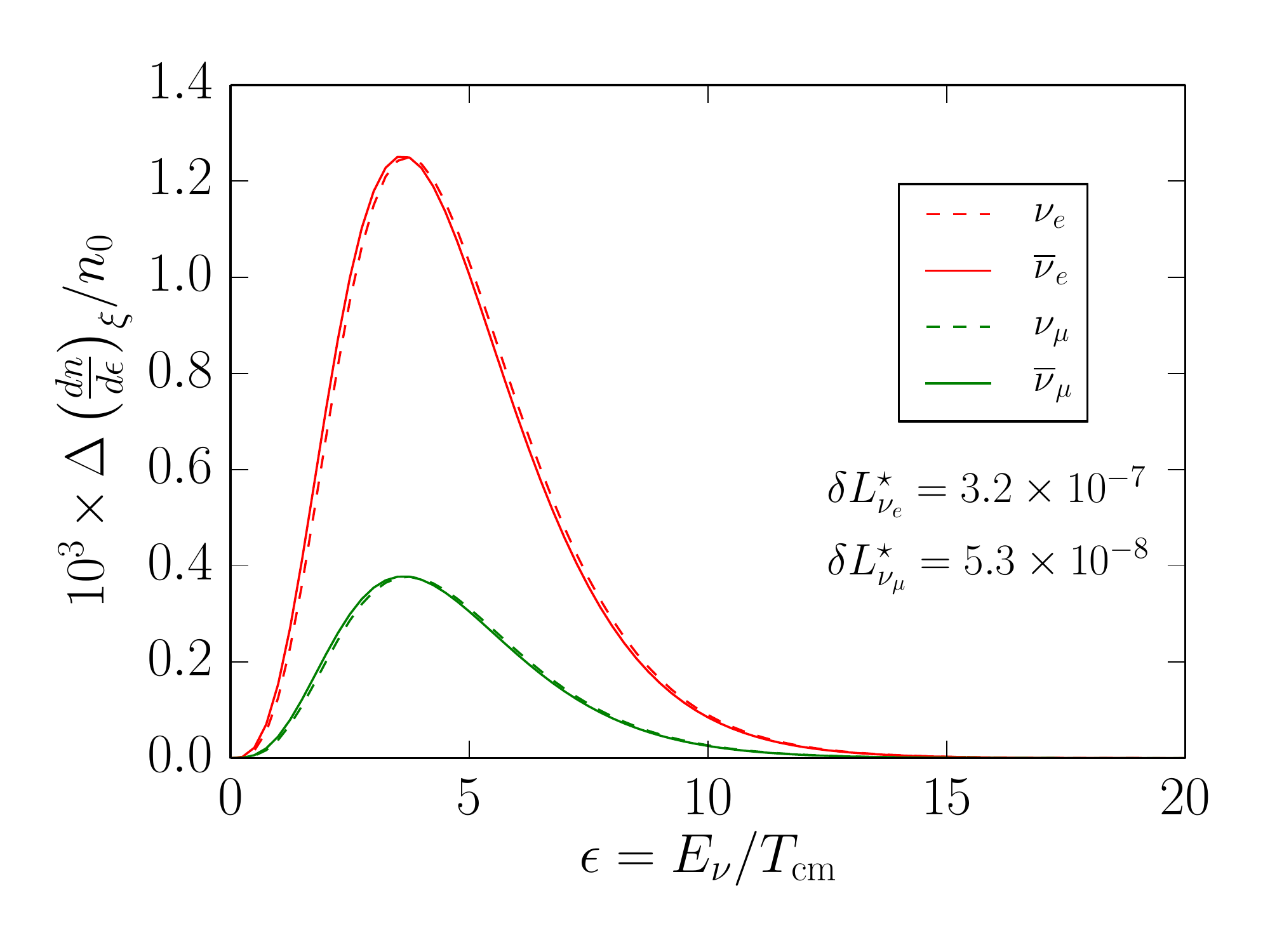}
   \end{center}
   \caption{\label{fig:num_annih}Absolute change in the neutrino/antineutrino
   number distributions plotted against \eps at $\tcm=1.0\,{\rm keV}$.  The
   numbers given on the plot ($\delta \lstari$ for $i=e,\mu$) show the relative
   error accumulated over the course of a simulation.
   }
\end{figure}

\begin{figure}
   \begin{center}
      \includegraphics[width=\columnwidth]{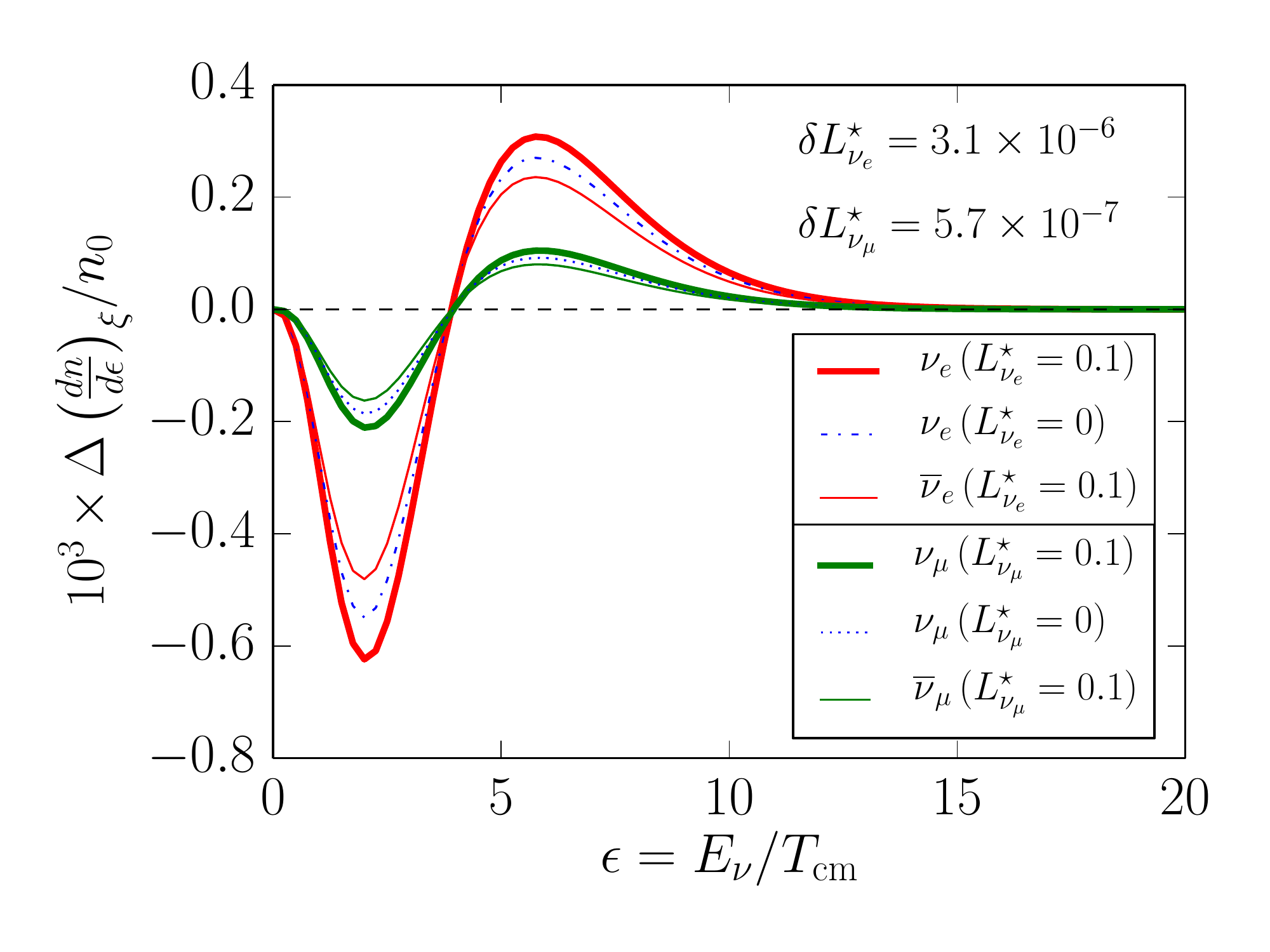}
   \end{center}
   \caption{\label{fig:num_nue}Absolute change in the neutrino/antineutrino
   number distributions plotted against \eps at $\tcm=1.0\,{\rm keV}$.  The
   numbers given on the plot ($\delta \lstari$ for $i=e,\mu$) show the relative
   error accumulated over the course of a simulation.
   }
\end{figure}

Figs.\ \ref{fig:num_annih} and \ref{fig:num_nue} show the absolute changes in
the number density distribution versus \eps when we include only certain
transport processes.  Fig.\ \ref{fig:num_annih} contains three annihilation
processes, schematically shown as:
\beq\label{rxn:annih}
  \nu_i + \bnu_i \leftrightarrow e^- + e^+,\quad i=e,\mu,\tau.
\eeq
In this scenario, we have included only the annihilation channel into
electron/positron pairs when computing transport.  The changes are with respect
to the same degeneracy parameters as those in Fig.\ \ref{fig:occ_xi_eps}.  The
line colors in Fig.\ \ref{fig:num_annih} correspond to the same species as
Fig.\ \ref{fig:occ_xi_eps}.  Because of the close proximity of the neutrino and
antineutrino curves, we depart from the previous nomenclature of emphasizing
the $\nu$ curves with a thicker line width so as not to obscure the \bnu curves.
For this plot, the absolute differences are normalized with respect to the
equilibrium number density at temperature \tcm with degeneracy parameter
$\xi=0$.  For a given neutrino species, the total change in number density
should be equal to the change in number density for the corresponding
antineutrino.

Fig.\ \ref{fig:num_nue} shows the effect of including 12 elastic scattering processes:
\begin{align}
  \nu_i + e^- &\leftrightarrow \nu_i + e^-,\label{rxn:nue1}\\
  \nu_i + e^+ &\leftrightarrow \nu_i + e^+,\label{rxn:nue2}
\end{align}
and the opposite-$CP$ reactions, for neutrino flavors $i=e,\mu,\tau$.  In this
scenario, we have included only the elastic scattering channel with
electrons/positrons (while neglecting the neutrino-antineutrino only channels)
when computing transport.  The changes are with respect to the same degeneracy
parameters as those in Fig.\ \ref{fig:occ_xi_eps}.  Furthermore, the line
colors and styles in Fig.\ \ref{fig:num_nue} correspond to the same species and
scenarios as Fig.\ \ref{fig:occ_xi_eps}.  For this plot, the absolute
differences are normalized with respect to the equilibrium number density at
temperature \tcm with degeneracy parameter $\xi=0$.  In an identical manner to
the processes in Fig.\ \ref{fig:num_annih}, the total change in $\nu$ number
density should be equal to the change in number density for the corresponding
\bnu in Fig.\ \ref{fig:num_nue}.

The elastic scattering processes of Eqs.\ \eqref{rxn:nue1} and \eqref{rxn:nue2}
(and the opposite-$CP$ reactions) preserve the total number of neutrinos and
antineutrinos.  The plasma of charged leptons acts to upscatter low energy
neutrinos and antineutrinos to higher energies, precipitating an entropy flow.
Fig.\ \ref{fig:num_nue} vividly shows a deficit of neutrinos in the range
$0<\eps\lesssim4$, and the corresponding excess for $\eps\gtrsim4$.  The
deficit is more pronounced in Fig.\ \ref{fig:num_nue} but also appeared in Figs.\
\ref{fig:occ_xi_eps}, \ref{fig:rho_xi_eps}, and \ref{fig:num_0_eps} when computing
the entire neutrino-transport network.  The annihilation processes, shown in
Fig.\ \ref{fig:num_annih}, do not preserve the total numbers of neutrinos and
antineutrinos and can fill the phase space vacated by the upscattered
neutrinos.  The complete transport network, which includes annihilation,
elastic scattering on charged leptons, and elastic scattering among only
neutrinos/antineutrinos, is able to redistribute the added energy by filling
the occupation numbers for lower epsilon.

\section{Integrated asymmetry measures}
\label{sec:asymm}

In our presentation to this point, we have used the comoving lepton number to
describe the asymmetry in the early universe.  \lstari does not evolve with
temperature in our model, except for errors in precision encountered by our
code.  Therefore, we introduce two integrated quantities to examine how the
initial asymmetry propagates to later times.  The quantities provide new means
to analyze the out-of-equilibrium spectra.

The first integrated quantity we define is the lepton energy density asymmetry
\beq\label{eq:R_def}
  R_i\equiv\frac{\rho_{\nu_i} - \rho_{\bnu_i}}{\frac{\pi^2}{15}\tcm^4}.
\eeq
where $i$ is the flavor index.  Like the comoving lepton number in Eq.\
\eqref{eq:lnu2}, we divide Eq.\ \eqref{eq:R_def} by $\tcm^4$ so that $R_i$ is
comoving and dimensionless.  This will allow us to follow the evolution of $R_i$ to later
times.  At large \tcm, all flavors have identical equilibrium FD spectra and
lepton numbers/degeneracy parameters.  For degeneracy parameter $\xi$, we
calculate the equilibrium value of $R$
\begin{align}
  R^{\rm (eq)} = {\rm sgn}(\xi)\left[\vphantom{\frac{7}{8}}\right.&\frac{7}{8}
  + \frac{15}{4}\left(\frac{\xi}{\pi}\right)^2
  + \frac{15}{8}\left(\frac{\xi}{\pi}\right)^4\nonumber\\
  -&\left. \frac{90}{\pi^4}e^{-|\xi|}\Phi(-e^{-|\xi|},4,1)\right],
\end{align}
where ${\rm sgn}(x)$ is the sign function with real-number argument $x$, and
$\Phi(z,s,v)$ is the Lerch function (see Sec.\ 9.55 of Ref.\
\cite{gradshteyn2007})
\beq
  \Phi(z,s,v)\equiv\sum\limits_{n=0}^{\infty}\frac{z^n}{(n+v)^s};
  \,|z|<1;\,v\ne0,-1,-2,....
\eeq

\begin{figure}
   \begin{center}
      \includegraphics[width=\columnwidth]{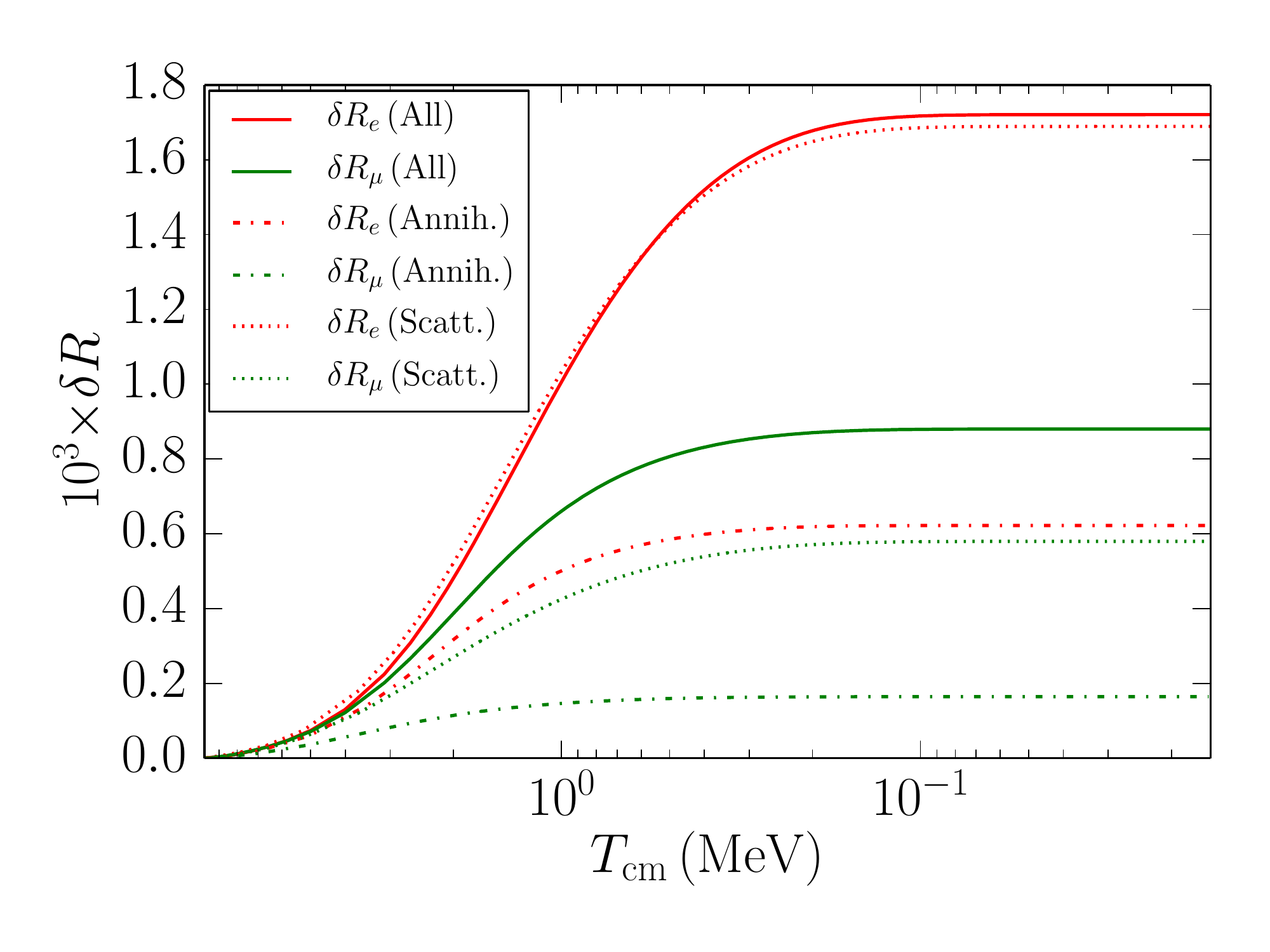}
   \end{center}
   \caption{\label{fig:dR_tcm_procs} The relative changes in $R$ plotted
   against \tcm for \nue and \num with different processes included in the
   transport calculation.  Red lines correspond to \nue and green lines correspond
   to \num.  The process scheme is all processes (solid curves), annihilation only
   (dash-dot curves), or elastic scattering only (dotted curves).
   }
\end{figure}
  
Figure \ref{fig:dR_tcm_procs} shows the relative changes in $R_i$ from the
$R^{\rm (eq)}$ baseline ($\delta R_i$), plotted against \tcm for different
combinations of transport processes.  Solid lines (All) are for the complete
calculation, whereas dash-dot curves only include the annihilation channels
(Annih.) of the reaction shown in \eqref{rxn:annih}, and dotted curves only
include the elastic scattering channels (Scatt.) of the reactions shown in
\eqref{rxn:nue1}, \eqref{rxn:nue2}, and the opposite-$CP$ reactions.  Red lines
correspond to $\delta R_e$ and green lines to $\delta R_\mu$.  $\delta R_i$
increases for all six combinations of flavor and transport process, until an
eventual freeze-out.  Indirectly, Figs.\ \ref{fig:num_0_eps},
\ref{fig:num_annih}, and \ref{fig:num_nue} all show that the neutrinos have
larger changes in the energy density distributions, increasing the asymmetry.
Because of the charged-current process, $\delta R_e$ experiences a greater
enhancement.  What is important to note is that the total $\delta R_i$, for
either flavor, is not an incoherent sum of the two transport processes taken
individually.  There are two reasons for this.

First, there are other transport processes in the full calculation.  Neutrinos
scattering on other neutrinos and antineutrinos will redistribute energy
density.  Second, the transport processes with the charged leptons are
dependent on one another.  Positron-electron annihilation into
neutrino-antineutrino pairs populates the lower energy levels.  Those particles
upscatter on charged leptons through elastic scattering.  Positron-electron
annihilation is then suppressed by the Pauli blocking of the upscattered
particles.  Both reasons change the evolution of the total $R_i$, but do so in
a flavor-dependent manner.  For $\delta R_\mu$, the incoherent sum of
annihilation and elastic scattering is smaller than that of the total
asymmetry.  For $\delta R_e$, the total asymmetry is dominated by the
contribution from elastic scattering.

\begin{figure}
   \begin{center}
      \includegraphics[width=\columnwidth]{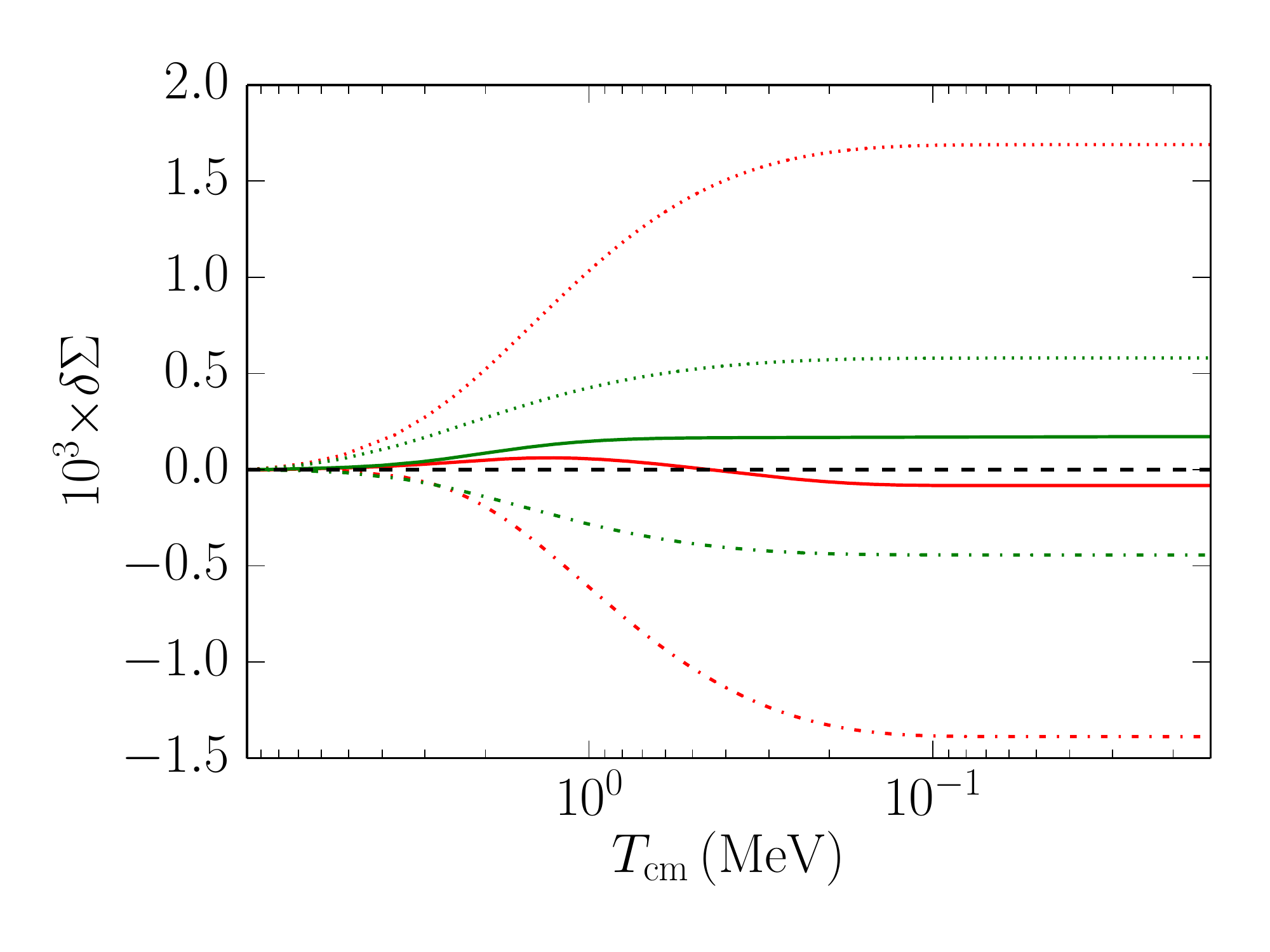}
   \end{center}
   \caption{\label{fig:dSig_tcm_procs} The relative changes in $\Sigma$ plotted
   against \tcm.  Line colors and styles correspond to the same transport
   processes and neutrino flavors in Fig.\ \ref{fig:dR_tcm_procs}.
   }
\end{figure}

In analogy with the lepton energy density asymmetry, we define the lepton
entropy asymmetry as
\beq
  \Sigma_i \equiv\frac{S_{\nu_i} - S_{\bnu_i}}{\frac{4}{3}\frac{\pi^2}{15}\tcm^3}
\eeq
where $S_j$ is the entropic density for particle $j$, given by
\beq\label{eq:s_def}
  S_j = -\frac{\tcm^3}{2\pi^2}\int_0^{\infty}d\eps\,\eps^2[f_j\ln f_j + (1-f_j)\ln(1-f_j)],
\eeq
and we have suppressed the arguments of $f_j(\eps;\xi)$ for brevity in
notation.  Under the equilibrium assumptions, we find
\beq
  \Sigma^{\rm (eq)} = R^{\rm (eq)} - \frac{45}{2\pi^4}\xi\left[\zeta(3)|L^{\star}_\nu|
  + e^{-|\xi|}\Phi(-e^{-|\xi|},3,1)\right].
\eeq
Fig.\ \ref{fig:dSig_tcm_procs} shows the evolution of the relative change in
$\Sigma_i$ away from $\Sigma^{\rm (eq)}$ when divided into processes.  The
nomenclature for the line styles and colors is identical to that in Fig.\
\ref{fig:dR_tcm_procs}.  The evolution of the lepton entropy asymmetry shows
more features than that of the lepton energy density asymmetry.

To understand the dynamics of $\Sigma_i$ in Fig.\ \ref{fig:dSig_tcm_procs}, we
begin by considering how the entropy depends on perturbations to the occupation
numbers.  We write the occupation numbers as differences from FD equilibrium
\beq\label{eq:f_pert}
  f_j(\eps;\xi) = \feq_j(\eps;\xi) + \Delta f_j(\eps;\xi).
\eeq
We can calculate the change in the entropy produced by the out-of-equilibrium
occupation numbers by substituting Eq.\ \eqref{eq:f_pert} into Eq.\
\eqref{eq:s_def}.  After dropping the \eps argument, $\xi$ argument, and species index
for notational brevity, we find for small $\Delta f$
\begin{widetext}
\begin{align}
  S &= -\frac{\tcm^3}{2\pi^2}\int_0^{\infty}d\eps\,
  \eps^2[(\feq+\Delta f)\ln(\feq+\Delta f) + (1-\feq-\Delta f)\ln(1-\feq-\Delta f)]\\
  &\simeq -\frac{\tcm^3}{2\pi^2}\int_0^{\infty}d\eps\,
  \eps^2\left[\feq\ln\feq + (1-\feq)\ln(1-\feq)+\Delta f\ln\frac{\feq}{1-\feq}\right]\\
  &= S^{\rm (eq)}-\frac{\tcm^3}{2\pi^2}\int_0^{\infty}d\eps\,
  \eps^2\Delta f[\xi-\eps]\\
  &= S^{\rm (eq)}-\xi\Delta n + \Delta\rho/\tcm,
\end{align}
\end{widetext}
where $\Delta n$ and $\Delta\rho$ are the changes in number and energy density,
respectively, from equilibrium.  The expression for the lepton entropy
asymmetry is
\beq\label{eq:Sigma1}
  \Sigma_i = \Sigma^{\rm (eq)} + \frac{45}{4\pi^2\tcm^3}
  \left[-\xi(\Delta n_{\nui} + \Delta n_{\bnui})
  + \frac{\Delta\rho_{\nui} - \Delta\rho_{\bnui}}{\tcm}\right].
\eeq
Lepton number is conserved in our scenarios, implying $\Delta n_{\nui} = \Delta
n_{\bnui}$.  As a result, we can write the lepton entropy asymmetry as
\beq\label{eq:Sigma2}
  \Sigma_i = \Sigma^{\rm (eq)} + \frac{45}{4\pi^2\tcm^3}
  \left[-2\xi\Delta n_{\nui} + \frac{\Delta\rho_{\nui} - \Delta\rho_{\bnui}}{\tcm}\right].
\eeq
Eq.\ \eqref{eq:Sigma2} shows how the lepton entropy asymmetry changes for small
perturbations to the occupation numbers.  Two trends are evident from this
equation.  First, adding particles ($\Delta n_{\nui}>0$) decreases the
asymmetry.  Second, increasing the asymmetry in energy density
($\Delta\rho_{\nui}-\Delta\rho_{\bnui}>0$), leads to an increase in the lepton
entropy asymmetry.  For the annihilation processes, the changes in the number
density distribution for neutrinos and antineutrinos vary in the same way
across \eps space for all flavors (see Fig.\ \ref{fig:num_annih}).  Therefore,
the corresponding changes in the energy density will also be the same, and
there will be no contribution to the change in $\Sigma$ from the energy density
terms.  The dash-dot curves in Fig.\ \ref{fig:dSig_tcm_procs} shows the
relative change in $\Sigma$ for a run with only the annihilation channels
active.  Both the $e$ and $\mu$ flavors show a suppression in $\Sigma$ with
decreasing \tcm.  Figure \ref{fig:num_nue} shows that for elastic scattering of
neutrinos and charged leptons, the neutrino and antineutrino number density
distributions are not coincident.  Overall, each neutrino species has zero net
change in number density, as elastic scattering can only redistribute the number.
Therefore, there will be no contribution to the change in $\Sigma$ from the
number density term.  As there are more neutrinos over antineutrinos for
$\lstarnu>0$, elastic scattering enhances the neutrino spectra over the
antineutrino spectra.  The result is a net positive change in the energy
density differences.  Fig.\ \ref{fig:dSig_tcm_procs} shows an increase in the
relative change in $\Sigma$ for the elastic-scattering-only runs for both
flavors.  When we add the elastic-scattering and annihilation channels
together, along with the other transport processes which do not involve charged
leptons, we see that the two processes essentially cancel, leaving only a
modest change in $\Sigma_i$ as shown by the solid lines in Fig.\
\ref{fig:dSig_tcm_procs}.

\begin{figure}
   \begin{center}
      \includegraphics[width=\columnwidth]{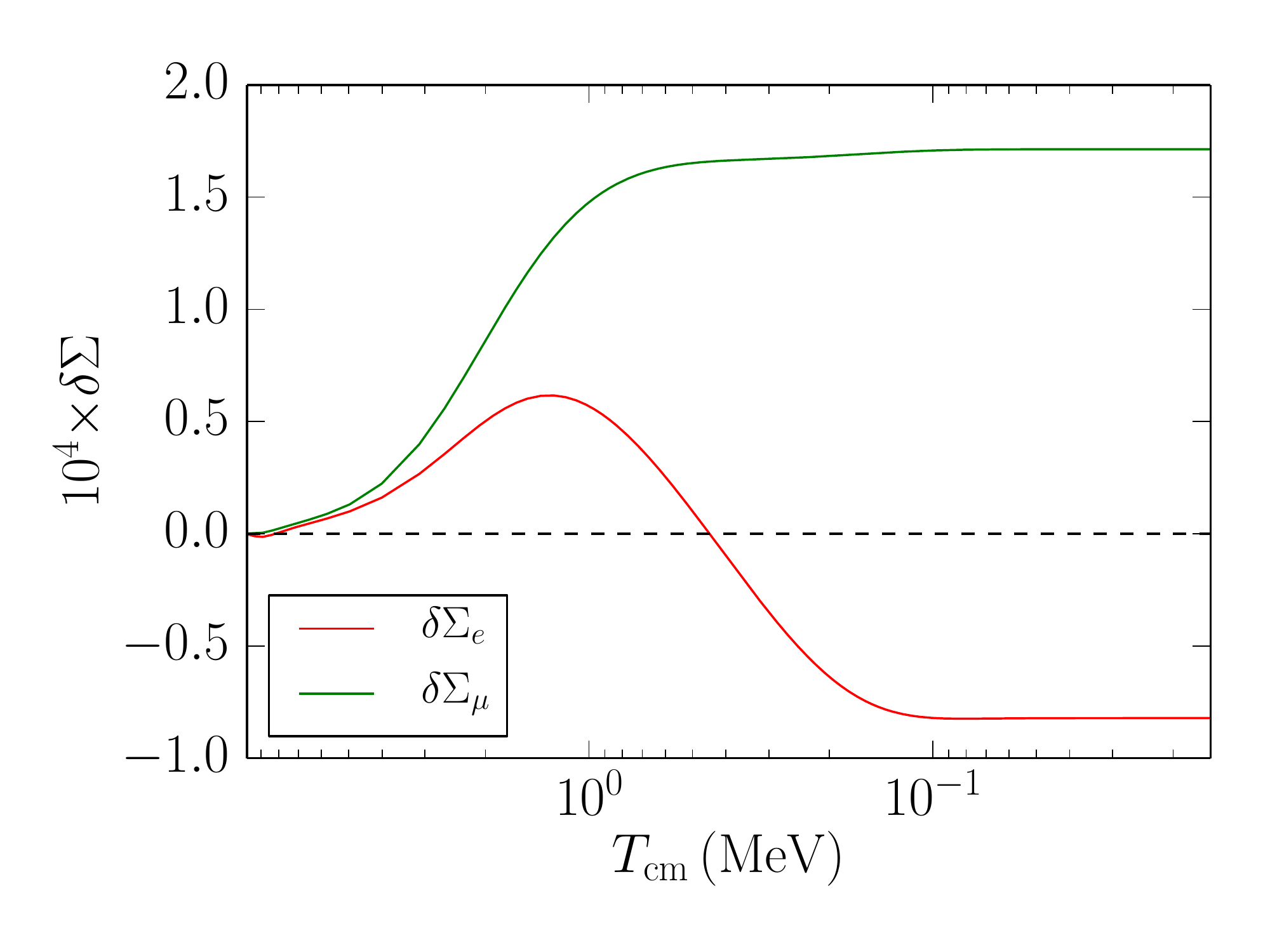}
   \end{center}
   \caption{\label{fig:dSig_tcm_all}Same as Fig.\ \ref{fig:dSig_tcm_procs}
   except zoomed in on the solid curves.
   }
\end{figure}

The interesting thing to note in Fig.\ \ref{fig:dSig_tcm_procs} is the
asymmetry between flavors.  Fig.\ \ref{fig:dSig_tcm_all} is a zoomed-in version
of the solid lines in Fig.\ \ref{fig:dSig_tcm_procs}.  We see that
$\delta\Sigma_\mu$ is monotonically increasing for decreasing \tcm.  The
incoherent sum of the relative changes from the annihilation and
elastic-scattering processes in Fig.\ \ref{fig:dSig_tcm_procs} nearly gives the
relative change in $\Sigma_\mu$ that we obtain when all transport processes are
active.  The same cannot be said for $\delta\Sigma_e$.  The sum of the two
transport processes is not incoherent, the evolution of $\delta\Sigma_e$ is not
monotonic, and the final freeze-out value of $\delta\Sigma_e$ is of opposite
sign from $\delta\Sigma_\mu$.  Although the elastic scattering would appear to
produce a larger enhancement of $\delta\Sigma_e$ over the suppression of
annihilation, the two processes do not have equal weight.  We observe this by
looking at the maxima in the number density distributions in Figs.\
\ref{fig:num_annih} and \ref{fig:num_nue}.  The ratio of maxima in Fig.\
\ref{fig:num_annih} for annihilation is
\beq
  \left(\frac{dn_{\nue}}{d\epsilon}\right)\biggr/
  \left(\frac{dn_{\num}}{d\epsilon}\right)\simeq3.5\quad{\rm (Annih.)}
\eeq
The ratio of maxima in Fig.\ \ref{fig:num_nue} for elastic scattering is
\beq
  \left(\frac{dn_{\nue}}{d\epsilon}\right)\biggr/
  \left(\frac{dn_{\num}}{d\epsilon}\right)\simeq3.0\quad{\rm (Scatt.)}
\eeq
This shows that annihilation is more dominant in the electron
neutrino/antineutrino sector than it is in the $\mu$ sector.  In Figs.\
\ref{fig:num_annih} and \ref{fig:num_nue}, we have only showed the final
distributions at freeze-out.  Electron-positron annihilation into neutrinos is
not always so dominant, as evidenced by the positive values of $\Sigma_e$ for
$\tcm\gtrsim400\,{\rm keV}$.

The analysis of the lepton entropy asymmetry focused on the transport processes
which involve the charged leptons.  The other scattering processes redistribute
occupation number and therefore change $\Sigma_i$.  However, we have verified
that the contributions from the transport processes which involve only
neutrinos or antineutrinos do not alter $\Sigma_i$ enough to explain the full
evolution shown in Fig.\ \ref{fig:dSig_tcm_all}.  The transport processes which
involve the charged leptons play the dominant roles.

We have considered the evolution of the integrated asymmetry measures for
$\lstarnu=0.1$ only.  Table \ref{tab:asymm} gives the relative changes in $R_i$
and $\Sigma_i$ at freeze-out for various values of \lstarnu.  Note that the
positive relative changes for $\lstarnu<0$ imply an absolute decrease in either
quantity.  We see that the differences between the various values of \lstarnu
are beneath the error floor.

\begin{table*}
  \begin{center}
  \begin{tabular}{| c !{\vrule width 1.5 pt} c | c !{\vrule width 1.5 pt} c | c |}
    \hline
    \lstarnu & $10^4\times\delta R_e$ & $10^4\times\delta R_\mu$ &
    $10^4\times\delta \Sigma_e$ & $10^4\times\delta \Sigma_\mu$ \\
    \midrule[1.5pt]
    $10^{-1}$ & $17.20$ & $8.802$ & $-0.8211$ & $1.714$ \\ \hline
    $10^{-2}$ & $17.26$ & $8.825$ & $-0.8804$ & $1.692$ \\ \hline
    $10^{-3}$ & $17.27$ & $8.828$ & $-0.8729$ & $1.694$ \\
    \midrule[1.5pt]
    $-10^{-1}$ & $17.20$ & $8.802$ & $-0.8214$ & $1.714$ \\ \hline
    $-10^{-2}$ & $17.26$ & $8.825$ & $-0.8820$ & $1.691$ \\ \hline
    $-10^{-3}$ & $17.24$ & $8.828$ & $-0.9001$ & $1.694$ \\ \hline
  \end{tabular}
  \end{center}
  \caption{\label{tab:asymm}Relative changes in integrated asymmetry measures
  at freeze-out for select values of comoving lepton number \lstarnu.
  }
\end{table*}

\section{Abundances}
\label{sec:abundances}

Our calculations show potentially significant changes in lepton-asymmetric BBN
abundance yields with neutrino transport relative to those without. With the
inclusion of transport we find that the general trends of the yields of \heiv
and D with increasing or decreasing lepton number are preserved: positive
\lstarnu decreasing the yields of both, while negative lepton numbers increase
both. In broad brush, Boltzmann transport makes little difference for helium,
but gives a $\ge0.3\%$ {\it reduction} in the offset from the FD, zero
lepton-number case with transport. This change in the reduction is comparable
to uncertainties in BBN calculations arising from nuclear cross sections and
from plasma physics and QED issues.  For all BBN calculations, the baryon to
photon ratio is fixed to be $n_b/n_\gamma=6.0747\times10^{-10}$ (equivalent to
the baryon density $\bardens=0.022068$ given by Ref.\ \cite{PlanckXVI:2014}).
In addition, the mean neutron lifetime is taken to be $885.7\,{\rm s}$.

Table \ref{tab:no_trans} contains relative differences in the primordial
abundances with and without transport.  Columns with the label ``FD Eq.'' are
the calculations without any active transport processes.  The spectra
freeze-out at high temperatures where they are in FD equilibrium with a
degeneracy parameter corresponding to \lstarnu.  Columns with the label
``Boltz.'' are the calculations in the full Boltzmann neutrino-transport
calculation.  Relative differences are with respect to the appropriate
abundance in the zero-degeneracy Boltz.\ calculation.  The relative changes in
the abundances for the two different calculations are quite close: $\delta\yp$
differs by 2 - 3 parts in $10^4$; and $\delta{\rm D/H}$ differs by 3 - 4 parts
in $10^3$.  Both differences are consistent across \lstarnu.  We caution
against any interpretation that links the two calculations together, as the No
Trans.\ calculations ignore important physics related to non-FD spectra,
entropy flow, and the Hubble expansion rate.

We have examined the detailed evolution of the
spectra and integrated asymmetry measures in the Boltz.\ calculations.  The electron neutrinos and
antineutrinos behave differently compared to muon and tau flavored neutrinos.
This behavior will have ramifications for the neutron-to-proton ratio and
nucleosynthesis.  To facilitate the analysis of the effects of neutrino
transport on BBN, we will introduce a model which uses additional radiation
energy density.  We will try to determine whether this simplistic ``dark
radiation'' model \cite{GFKP-5pts:2014mn,2000PhLB..473..241M} -- which includes
radiation energy density distinct from photons and active neutrinos, but does
not include transport -- can mock up the effects of the extra energy density
which arise from neutrino scattering and the associated spectral distortions.
We will compare this dark-radiation model to the full neutrino-transport case.
For ease in notation when comparing the two scenarios, we will abbreviate the
dark-radiation model as ``DR'' and the full Boltzmann neutrino-transport
calculation as Boltz.

\begin{table*}
  \begin{center}
  \begin{tabular}{| c !{\vrule width 1.5 pt} c | c
  !{\vrule width 1.5 pt} c | c |}
    \hline
    $\lstarnu$ 
    & $\delta Y_P$ (FD Eq.) & $\delta Y_P$ (Boltz.)
    & $\delta({\rm D/H})$ (FD Eq.) & $\delta({\rm D/H})$ (Boltz.) \\ 
    \midrule[1.5pt]
    $10^{-1}$ & $-0.1333$ & $-0.1331$
    & $-6.972\times10^{-2}$ & $-6.654\times10^{-2}$\\ \hline
    $10^{-2}$ & $-1.425\times10^{-2}$ & $-1.400\times10^{-2}$
    & $-1.101\times10^{-2}$ & $-7.618\times10^{-3}$\\ \hline
    $10^{-3}$ & $-1.678\times10^{-3}$ & $-1.409\times10^{-3}$
    & $-4.206\times10^{-3}$ & $-7.815\times10^{-4}$ \\ \hline
    $7.139\times10^{-3}$ & $-1.027\times10^{-2}$ & $-1.001\times10^{-2}$
    & $-8.867\times10^{-3}$ & $-5.463\times10^{-3}$ \\ \hline
    $1.364\times10^{-2}$ & $-1.931\times10^{-2}$ & $-1.906\times10^{-2}$
    & $-1.371\times10^{-2}$ & $-1.033\times10^{-2}$ \\
    \midrule[1.5pt]
    $-10^{-1}$ & $0.1475$ & $0.1479$
    & $8.566\times10^{-2}$ & $8.947\times10^{-2}$\\ \hline
    $-10^{-2}$ & $1.384\times10^{-2}$ & $1.415\times10^{-2}$
    & $4.352\times10^{-3}$ & $7.825\times10^{-3}$\\ \hline
    $-10^{-3}$ & $1.133\times10^{-3}$ & $1.407\times10^{-3}$
    & $-2.669\times10^{-3}$ & $7.630\times10^{-4}$\\ \hline
    $-7.071\times10^{-3}$ & $9.692\times10^{-3}$ & $9.968\times10^{-3}$
    & $2.047\times10^{-3}$ & $5.495\times10^{-3}$\\ \hline
    $-1.240\times10^{-2}$ & $1.724\times10^{-2}$ & $1.756\times10^{-2}$
    & $6.253\times10^{-3}$ & $9.740\times10^{-3}$\\ \hline
  \end{tabular}
  \end{center}
  \caption{\label{tab:no_trans}Relative changes in primordial abundances of
  $^4{\rm He}$ and D in two calculations of BBN with nonzero comoving lepton
  numbers \lstarnu.  FD Eq.\ signifies the calculation without transport.
  Boltz.\ signifies the full Boltzmann neutrino-transport network calculation.
  The abundances are given as relative changes from the zero degeneracy, full
  Boltzmann calculation.  Column 1 is the comoving lepton number.  Column 2
  gives the relative change of $Y_P$ at freeze-out in the no-transport model.
  Column 3 gives the relative change of $Y_P$ at freeze-out in the Boltz.\
  calculation.  The relative changes for D/H are given in columns 4 and 5.  The
  four rows where $|\lstarnu|$ is not a power of 10 are projected sensitivity limits
  for $1\%$ changes in the primordial abundances.
  }
\end{table*}


In the DR model, we introduce extra radiation energy density, $\rho_{\rm dr}$,
described at early times by the dark-radiation parameter \deltadr
\beq\label{eq:dr}
  \rho_{\rm dr} = \frac{7}{8}\frac{\pi^2}{15}\deltadr\tcm^4.
\eeq
The FD Eq.\ calculation in Table \ref{tab:no_trans} used $\deltadr=0$.  We
mandate that the dark radiation be composed of relativistic particles which are
not active neutrinos.  We have chosen the specific form of Eq.\ \eqref{eq:dr}
for conformity with \neff, namely $\Delta\neff\approx\deltadr$.  The relation
is not a strict equality due to the presence of finite-temperature-QED
corrections to the electron rest mass
\cite{transport_paper,1994PhRvD..49..611H,1997PhRvD..56.5123F,
1982NuPhB.209..372C,1999PhRvD..59j3502L}.  The DR model differs from the
Boltz.\ calculation in multiple respects.  First, the DR model fixes the
neutrino spectra to be in degenerate FD equilibrium.  Second, neutrino
transport induces an entropy flow from the plasma into the neutrino seas,
absent in the DR model.  Third, the entropy flow changes the phasing of the
plasma temperature with the comoving temperature parameter as compared to the
case of instantaneous neutrino decoupling in the DR model.  The phasing is
dependent on the Hubble expansion rate and the flow of entropy.  Although the
expansion rates are identical in the two scenarios, the entropy flows are not.

For all calculations, we will fix $\deltadr = 0.03149$.  We pick this specific
value to match \neff between the DR model and Boltz.\ calculation for the single case
$\lstarnu=0.1$.  The change in \neff depends on the Hubble expansion rate,
which depends on the initial degeneracy.  Therefore, our choice of \deltadr
will not ensure equal values of \neff between the two scenarios for
$\lstarnu\ne0.1$.  Although our DR model is not consistent across all \lstarnu,
the changes in \neff are small for the range of \lstarnu we explored.

\begin{figure}
   \begin{center}
      \includegraphics[width=\columnwidth]{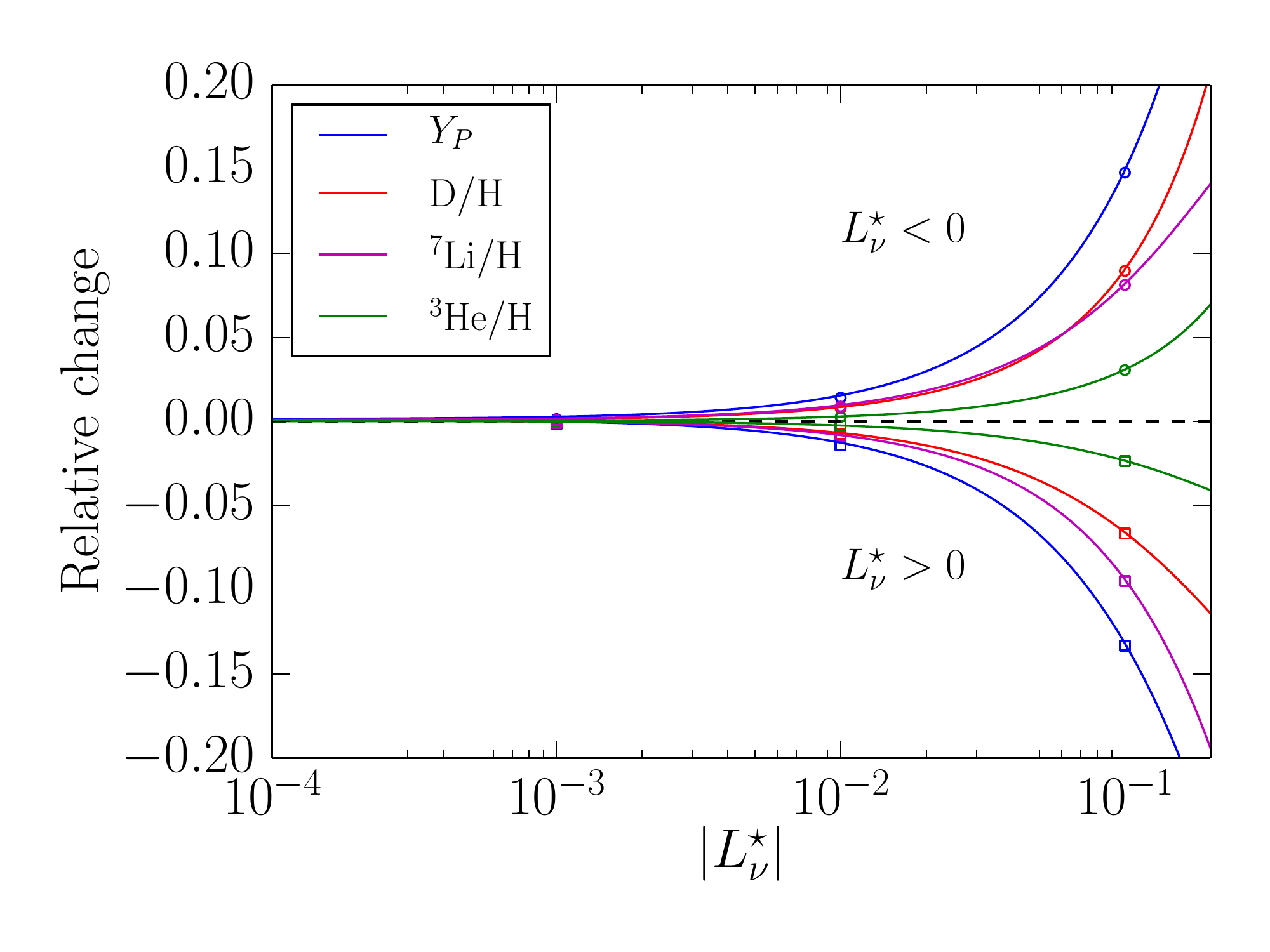}
   \end{center}
   \caption{\label{fig:abunds_v_lep}Relative changes in the primordial
   abundances plotted against the absolute value of the comoving lepton number.
   Positive changes in abundances correspond to negative comoving lepton numbers,
   and negative changes correspond to positive comoving lepton numbers.  The solid
   lines use the dark-radiation model described in the text.  Individual points
   using the full Boltzmann-transport calculation are plotted for three decades of
   \lstarnu.  Squares correspond to $\lstarnu>0$, and circles for $\lstarnu<0$.
   The baryon density is fixed to be $\bardens=0.022068$ (equivalent to a
   baryon-to-photon ratio $n_b/n_\gamma=6.0747\times10^{-10}$) for all
   calculations in both scenarios \cite{PlanckXVI:2014}.  .  The mean neutron
   lifetime is taken to be $885.7\,{\rm s}$.
   }
\end{figure}

Figure \ref{fig:abunds_v_lep} shows the relative changes in abundances versus
the comoving lepton number for both calculations.  Our baselines for
comparison are the abundances in the nondegenerate case, $\lstarnu=0$, from the
Boltz.\ calculation.  As a result of the choice of baseline, the relative
changes in abundances for the DR model will not converge to zero as
$\lstarnu\rightarrow0$.  We use a mass fraction to describe the helium
abundance, \yp, and relative abundances with respect to hydrogen to describe
deuterium (D), helium-3 ($^3{\rm He}$), and lithium-7 ($^7{\rm Li}$).  The
solid lines in Fig.\ \ref{fig:abunds_v_lep} show the relative changes in the DR
model.  Positive relative changes in the abundances correspond to negative
comoving lepton numbers, and negative changes to positive \lstarnu.  We also
show individual points using the Boltz.\ calculation at three decades of
\lstarnu, namely $\log_{10}|\lstarnu|=-1, -2, -3$.  Squares correspond to
$\lstarnu>0$, and circles to $\lstarnu<0$.

All abundances decrease with increasing \lstarnu.  A nonzero comoving lepton
number changes the occupation numbers in the neutron-proton interconversion
rates, and also changes the Hubble expansion rate.  The neutron-to-proton ratio
($n/p$) is sensitive to both quantities
\cite{2009PhRvD..79j5001S,2016NuPhB.911..955G}, and \yp is the abundance most
sensitive to $n/p$.  In Fig.\ \ref{fig:abunds_v_lep}, we see that \yp has the
largest change from the nondegenerate baseline, while \threehe has the least
sensitivity to \lstarnu.  Deuterium and \livii have a more intricate
relationship with \lstarnu.  As we increase \lstarnu from large negative values
towards zero, we see that the relative change for D is larger than that for
\livii until $\lstarnu\sim-5\times10^{-2}$.  At this point, \livii appears to
be more sensitive to \lstarnu.  The trend continues for $\lstarnu>0$, as the
relative change in \livii is more negative than that of D.  The asymmetry
between $\lstarnu>0$ and $\lstarnu<0$ in the relative changes of D and \livii
is present in \yp and \threehe also.  With the exception of \livii, all
abundances are more sensitive to negative \lstarnu.  All trends occur in both
the DR model and Boltz.\ calculation.  These trends are similar but have minor
differences than those discussed in Ref.\ \cite{2004NJPh....6..117K}.

Table \ref{tab:abundances} gives the relative changes of \yp and D/H for
various values of \lstarnu in both scenarios.  Columns with the label ``(DR)''
are relative changes calculated with the dark-radiation model and columns with
the label ``(Boltz.)'' are relative changes in the full Boltzmann
neutrino-transport calculation.  The Boltz.\ columns in Table
\ref{tab:abundances} are identical to the Boltz.\ columns in Table
\ref{tab:no_trans}.  For all four abundance columns, the relative changes are
with respect to the abundance calculated with the full Boltzmann-transport
network with degeneracy parameter set to zero, consistent with the lines and
points in Fig.\ \ref{fig:abunds_v_lep}.  For the Boltz.\ columns, the relative
changes in \yp tend to be twice as large as those in D/H.  Each decade change
in \lstarnu induces close to a decade change in both relative abundances.  We
have included calculations for sets of lepton numbers which aim for $\pm1\%$
changes in both \yp and D/H in the Boltz.\ calculation.  For the DR model, the
relative changes for \heiv and deuterium are in line with the Boltz.\
calculation for $\lstarnu=+0.1$.  Transport enhances the \nue occupation
numbers over the \bnue if $\lstarnu=+0.1$.  The extra probability in the \nue
spectrum enhances the rate of $\nue+n\rightarrow p+e^-$.  As a result, the
helium abundance decreases further in the Boltz.\ calculation, which is evident
in Table \ref{tab:abundances}.  Conversely, for $\lstarnu=-0.1$, transport will
enhance the \bnue over the \nue and we would expect an increase in the \heiv
abundance.  This is not the case in Table \ref{tab:abundances} - the DR model
has a larger $\delta\yp$ than the Boltz.\ calculation.

\begin{table*}
  \begin{center}
  \begin{tabular}{| c !{\vrule width 1.5 pt} c | c !{\vrule width 1.5 pt} c | c |}
    \hline
    \lstarnu & $\delta Y_P$ (DR) & $\delta Y_P$ (Boltz.)
    & $\delta {\rm (D/H)}$ (DR) & $\delta {\rm (D/H)}$ (Boltz.) \\
    \midrule[1.5pt]
    $10^{-1}$ & $-0.1318$ & $-0.1331$ & $-6.589\times10^{-2}$ & $-6.654\times10^{-2}$ \\ \hline
    $10^{-2}$ & $-1.257\times10^{-2}$ & $-1.400\times10^{-2}$
    & $-6.823\times10^{-3}$ & $-7.618\times10^{-3}$ \\ \hline
    $10^{-3}$ & $2.683\times10^{-5}$ & $-1.409\times10^{-3}$
    & $2.054\times10^{-5}$ & $-7.815\times10^{-4}$ \\ \hline
    $7.139\times10^{-3}$ & $-8.576\times10^{-3}$ & $-1.001\times10^{-2}$
    & $-4.467\times10^{-3}$ & $-5.463\times10^{-3}$ \\ \hline
    $1.364\times10^{-2}$ & $-1.762\times10^{-2}$ & $-1.906\times10^{-2}$
    & $-9.540\times10^{-3}$ & $-1.033\times10^{-2}$ \\
    \midrule[1.5pt]
    $-10^{-1}$ & $0.1494$ & $0.1479$ & $9.038\times10^{-2}$ & $8.947\times10^{-2}$ \\ \hline
    $-10^{-2}$ & $1.556\times10^{-2}$ & $1.415\times10^{-2}$
    & $8.627\times10^{-3}$ & $7.825\times10^{-3}$ \\ \hline
    $-10^{-3}$ & $2.840\times10^{-3}$ & $1.407\times10^{-3}$
    & $1.566\times10^{-3}$ & $7.630\times10^{-4}$ \\ \hline
    $-7.071\times10^{-3}$ & $1.141\times10^{-2}$ & $9.968\times10^{-3}$
    & $6.309\times10^{-3}$ & $5.495\times10^{-3}$ \\ \hline
    $-1.240\times10^{-2}$ & $1.897\times10^{-2}$ & $1.756\times10^{-2}$
    & $1.054\times10^{-2}$ & $9.740\times10^{-3}$ \\ \hline
  \end{tabular}
  \end{center}
  \caption{\label{tab:abundances}Relative changes in primordial abundances of
  $^4{\rm He}$ and D in two calculations of neutrino transport with nonzero
  comoving lepton numbers \lstarnu.  DR signifies the dark-radiation model of
  neutrino transport and Boltz.\ signifies the full Boltzmann neutrino-transport
  network calculation.  The columns are the same as Table \ref{tab:no_trans} with
  the replacement of ``FD Eq.'' by DR.
  }
\end{table*}

The error in the above logic resides in the treatment of the rate which changes
protons to neutrons, namely $\bnue+p\rightarrow n + e^+$.  This reaction has a
threshold of $Q\equiv\deltamnp+m_e\simeq1.8\,{\rm MeV}$, where \deltamnp and
$m_e$ are the neutron-to-proton mass difference and electron rest mass,
respectively.  If we define the appropriate \epsval for $Q$ to be $q\equiv
Q/\tcm$, we can see where and how the threshold plays a role in \eps and \tcm
space.  Fig.\ \ref{fig:num_0_eps} shows the freeze-out distortion to the
differential number density distributions for $\lstarnu=+0.1$.  The \nue and
\bnue spectra would be switched if we had plotted $\lstarnu=-0.1$, i.e.\ a
``mirror'' of Fig.\ \ref{fig:num_0_eps}.  At the start of the calculation at
$\tcm=10\,{\rm MeV}$, the distortions are identically zero.  The calculation
proceeds and the peaks in $\Delta(dn/d\epsilon)$ for \nue and \bnue grow.  The
locations of the peaks do change with decreasing \tcm, but we have verified
that the shift in position is small compared to peak location of $\eps\sim4$.
We claimed above that the extra number density of \bnue over \nue would
increase the rate $\bnue+p\rightarrow n+e^+$, but it is only the number density
with \epsval larger than $q$ which is able to increase the rate, thereby
decreasing the neutron abundance.  At $\tcm=1\,{\rm MeV}$, $q\simeq2$
which is large enough to exclude a portion of the left-hand side of the peak,
effectively limiting the number of antineutrinos which could participate in the
channel $\bnue+p\rightarrow n+e^+$.  At $\tcm=500\,{\rm keV}$, $q\simeq4$ which
is nearly coincident with the central location of the peak.  This is the point
in \tcm where the \heiv abundance begins to depart from nuclear statistical
equilibrium \cite{SMK:1993bb}.  Although the abundance is $\sim15$ orders of
magnitude smaller than its freeze-out value, the integration of the nuclear
reaction network is sensitive to the initial conditions, and already half of
the peak width in the mirror of Fig.\ \ref{fig:num_0_eps} is unavailable to
enhance the rate and modify the neutron-to-proton ratio.  The formation of
\heiv nuclei is typically ascribed to the epoch $\tcm=100\,{\rm keV}$, where
$q\simeq20$ and well larger than the range where the distortions in the mirror
of Fig.\ \ref{fig:num_0_eps} could affect the rate for $\bnue+p\rightarrow
n+e^+$.  Meanwhile, neutrino transport is inducing an increased population on
the high-energy tail of the \nue spectrum, which would increase the neutron to
proton rate $\nue+n\rightarrow p+e^-$.  This reaction has no threshold, and so
the entire peak in the mirror of Fig.\ \ref{fig:num_0_eps}, integrated over the
full range of \tcm, would increase the rate.  Incidentally, $e^++n\rightarrow
p+\bnue$ has no threshold and this process is also important in setting \np.
However, in this case, the spectral distortion effects we described above would
tend to hinder this process by producing extra \bnue blocking.  The \bnue in
$e^++n\rightarrow p+\bnue$ has a minimum energy of $Q$, and so the expected
suppression of this rate from additional \bnue number density suffers from the
same sequence of events as mentioned above.

To summarize, transport-induced \nue and \bnue spectral distortions develop
over such a long time span that the threshold-limited \bnue number density
cannot overcome the \nue number density when calculating the neutron-proton
interconversion rates in the $\lstarnu=-0.1$ case.  The result is a
decrease in $\delta\yp$ for the Boltz.\ calculation compared to the DR model.

The DR model is tuned to have the same total energy density as produced in the
full Boltzmann calculation when $\lstarnu=\pm0.1$.  If $|\lstarnu|\ne0.1$, the
radiation energy density, and by extension \neff, is slightly different.  The
abundances are sensitive to the change in \neff, and as a result we see
significant differences between the two models in Table \ref{tab:abundances}.
An especially egregious example is the $\lstarnu=10^{-3}$ scenario, where the
relative changes in \heiv are 2 orders of magnitude different and have
different signs.  We conclude that mocking up the effect of neutrino transport
in this model with dark radiation fails for small lepton numbers.  However, if
we had tuned the DR model for \neff to agree when $\lstarnu=10^{-3}$, we would
have had better agreement for smaller \lstarnu.  We note that for all cases
with $|\lstarnu|\le7\times10^{-3}$, the changes in the abundances are below
current and projected error tolerances \cite{cmbs4_science_book}.

\section{Conclusion}
\label{sec:concl}

We have done the first nonzero neutrino chemical potential calculations of weak
decoupling and BBN with full Boltzmann neutrino transport
simultaneously coupled with all relevant strong, weak, and electromagnetic
nuclear reactions. We have performed these calculations with a modified version
of the \burst code. This code and the physics it incorporates is described in
detail in Ref.\ \cite{transport_paper}. By design, our calculations here do not
include neutrino flavor oscillations. Our intent was to provide {\it baseline}
calculations for comparison to future neutrino flavor quantum kinetic
treatments (see Refs.\ \cite{1991NuPhB.349..743B,2013PhRvD..87k3010V} in the
early universe, and Refs.\ \cite{1993APh.....1..165R, 2007JPhG...34...47B,
2015IJMPE..2441009V} in core-collapse supernova cores, for a discussion on the
quantum kinetic equations in their respective environments). One objective of
this baseline Boltzmann study was to identify how a significant lepton number
would affect out-of-equilibrium neutrino scattering and the concomitant
neutrino scattering-induced flow of entropy out of the photon-electron-positron
plasma and into the decoupling neutrino component.  A related objective was to
assess whether (and how) the scattering-induced neutrino spectral distortions
develop differently in the case of a significant neutrino asymmetry.  The third
objective was to use a new description to connect the two previously mentioned
phenomena: macroscopic thermodynamics of entropy flow, and microscopic spectral
distortions.  Finally, the last objective was to assess the impact of these
neutrino spectral distortions and the accompanying changes in entropy flow and
temperature/scale factor phasing on BBN light element abundance yields.  A key
finding of our full Boltzmann neutrino-transport treatment is that the presence
of a lepton-number asymmetry {\it enhances} the processes which give rise to
distortions from equilibrium, FD-shaped neutrino and antineutrino energy
spectra. Our transport calculations show a positive feedback between
out-of-equilibrium neutrino scattering and any initial distortion from a zero
chemical potential FD distribution (see the elastic scattering of neutrinos
with charged leptons in Fig.\ \ref{fig:num_nue}). An initial distortion, for
example, stemming from a nonzero chemical potential, is amplified by neutrino
scattering, at least for higher values of the comoving neutrino energy parameter
$\epsilon = E_\nu/\tcm$. Of course, overall lepton asymmetry is preserved by
the nonlepton number violating scattering processes we treat here.

In broad brush, as the Universe expands entropy is transferred from the
electron-positron component into photons, with neutrinos receiving only a small
portion of this entropy largess. The magnitude of this small entropy increase
to the decoupling neutrinos is governed largely by the out-of-equilibrium
scattering of neutrinos and antineutrinos on the electrons and positrons, which
are generally ``hotter'' than the neutrinos.
The neutrino scattering cross sections scale like $\sigma\sim \epsilon^2$, and
therefore higher energy neutrinos are able to extract entropy from the
photon-electron-positron component more effectively than neutrinos with lower
energy.  The result is that a ``bump'' or occupation excess (see Fig.\
\ref{fig:num_0_eps}) on the higher energy end of the neutrino energy
distribution function grows with time.  Our transport calculations have allowed
us to track both entropy flow between the neutrinos and the plasma and the
simultaneous development of neutrino spectral distortions, all for a range of
initial lepton asymmetries. For the larger values of lepton asymmetry
considered here we found that the entropy transferred to neutrinos is decreased
by a few tenths of a percent over the zero lepton-number case (see Table
\ref{tab:lep_w_trans}). 

The enhanced neutrino spectral distortions and entropy transfer revealed by our
full Boltzmann-transport calculations might be expected to translate into
corresponding nuclear abundance changes emerging from BBN. Our full coupling
between neutrino scattering and the weak interaction sector and the nuclear
reaction network is uniquely adapted to treat this physics. Indeed, for the
zero neutrino chemical potential cases, the full Boltzmann-neutrino transport
resulted in a deuterium BBN yield $\sim 1\,\%$ different than a calculation
with no neutrino transport and a sharp weak decoupling approximation (see Table
V of Ref.\ \cite{transport_paper}). The baseline Boltzmann transport
calculations with significant lepton asymmetries reported here show that the
shift in BBN abundances with nonzero neutrino chemical potentials are closely
in line with those reported in sharp weak decoupling studies
\cite{2004NJPh....6..117K}, but with a few peculiarities. The enhanced spectral
distortions discussed above for the lepton asymmetry cases do alter the
charged-current weak interaction neutron-to-proton interconversion rates and,
in turn, this leads to altered abundance yields over the no-transport, sharp
decoupling treatment. To put these alterations in perspective, our full
Boltzmann calculations of BBN show that the \heiv abundance yield is sensitive
at the one percent level to an initial, comoving lepton number of
$\lstarnu\approx 7\times{10}^{-3}$, while the deuterium abundance yield is
similarly sensitive to $\lstarnu\approx 1.5\times{10}^{-2}$.  This is
significant because the next generation CMB experiments, e.g.\ proposed Stage-4
CMB observations \cite{cmbs4_science_book}, target precisions for independent
primordial helium abundance determinations at roughly the two percent level.
Likewise, the next generation of large optical telescopes, for example 30-meter
class telescopes \cite{TMT,2015RAA....15.1945S,GMT,EELT}, are touted as
providing a comparable level of precision in determining the primordial
deuterium abundance from quasar absorption lines in high redshift damped
Lyman-alpha systems. Our calculations show that we would need $\sim 0.1\,\%$
precision in these primordial abundance determinations to probe different
treatments of neutrino scattering in the weak decoupling epoch, at least for
the case with no neutrino oscillations. 

Though our calculations show that the bulk of the alteration in abundances
stems from the initial lepton asymmetry itself, transport does produce offsets
in absolute abundances yields comparable to those with zero lepton numbers. We
found that sometimes we can adequately capture the BBN effects of full
Boltzmann neutrino transport by using a dark radiation model of extra
radiation energy density added by neutrino scattering. However, this
approximation, tuned to agree with the Boltzmann calculation results at one
value of comoving lepton number, fails for other lepton asymmetry values.

We showed in Table \ref{tab:abundances} how neutrino transport alters the
primordial abundances in degenerate cases.  Both \heiv and D are sensitive to
\np, which itself is sensitive to the \nue and \bnue occupation numbers.  Table
\ref{tab:no_trans} showed that the FD Eq.\ treatment of BBN closely matches the
Boltz.\ calculation of \yp.  Transport induces a relative change in D/H nearly
an order or magnitude larger than that of \yp.  This finding is consistent with
findings in the zero-degeneracy case \cite{transport_paper}.  Tables
\ref{tab:lep_w_trans} and \ref{tab:abundances} show that the primordial
abundances are more sensitive to neutrino degeneracy than \neff.  Moreover,
\heiv is twice as sensitive to the degeneracy than D.  CMB Stage-IV experiments
\cite{cmbs4_science_book,Abazajian201566} and 30-meter-class telescopes will
probe \yp, D/H, and \neff at the $1\%$ level.  If future observations were to
find little change in \neff from the standard prediction, but changes in the
abundances matching the patterns in Table \ref{tab:abundances}, then this
scenario would be consistent with a degeneracy in the neutrino sector.
However, the Boltz.\ calculations in Table \ref{tab:abundances} do not include
the physics of neutrino oscillations.  In the presence of nonzero lepton
numbers, oscillations may alter the scaling relations of Table
\ref{tab:abundances} and will necessitate a full quantum kinetic equation
treatment \cite{VFC:QKE,Blaschke_Cirigliano_2016}.

This brings us to the question of our selection of initial lepton asymmetries.
We have chosen to examine values of these at and below usually accepted limits,
and we have examined only situations where the asymmetries are the same across
all flavors. The trends our Boltzmann-transport calculations reveal will likely
hold for lepton asymmetries outside of the ranges considered here. However,
differences in lepton numbers between different flavors will drive
medium-enhanced/affected neutrino flavor transformation which could lead to
different conclusions in the neutrino sector.  Comparing future quantum kinetic
calculations which include both coherent and scattering-induced flavor
transformation with our strict Boltzmann treatment might reveal BBN and \neff
signatures of neutrino flavor conversion, although these may be at levels well
below what future observations and experiments can probe.

Nevertheless, many beyond-standard-model physics considerations invoke
quite small initial lepton numbers
\cite{1990PhRvD..42.3344H,2002PhRvD..66d3516K,2010PhRvD..81h5032B,2014JCAP...03..028M}.
Various models of sterile neutrinos in the early universe, including dark
matter models, rely on lepton number-driven medium enhancements
\cite{1999PhRvL..82.2832S,AFP_2001,2006PhRvL..97n1301K} or beyond-standard-model physics to
create relic sterile-neutrino densities (see Refs.\
\cite{2013arXiv1310.4340D,2016arXiv160204816A} and references therein for a
review of sterile neutrino dark matter).  Sterile neutrinos are an intriguing
dark matter candidate \cite{Dod_Wid_1994}, and could conceivably be congruent
with particle \cite{2010PhLB..693..144K} and cosmological bounds
\cite{2003PhRvD..68f3507Y,2014PhRvL.112p1303A}.  For resonantly produced
sterile neutrino dark matter, the models invoke lepton asymmetries in the
${10}^{-3}$ to ${10}^{-5}$ range to match the relic dark-matter abundance,
providing a motivation for our choice of values for \lstarnu.

In fact, many models for baryon and lepton-number generation in the early
universe \cite{1999JHEP...11..015M,2010PhRvD..82i3009G,2013PhRvL.110f1801C} ,
e.g.\ the neutrino minimal standard model ($\nu$MSM)
\cite{2005PhLB..631..151A,2006PhLB..639..414S} , can produce lepton numbers in
the ranges chosen for the the present study.  It will be interesting to see if
future quantum kinetic calculations with neutrino flavor transformation will
yield deviations from the baseline calculations presented here. Any such
deviations would point to either a different distribution of lepton numbers
over neutrino flavor than that considered here, or differences in the
development of scattering-induced spectral distortions and attendant BBN
abundance alterations over the standard scenario.

\acknowledgments 
We thank Fred Adams, J.\ Richard Bond, Lauren Gilbert, Luke Johns, Joel Meyers,
Matthew Wilson, and Nicole Vassh for useful conversations.  We acknowledge the
Integrated Computing Network at Los Alamos National Laboratory for
supercomputer time.  This research used resources of the National Energy
Research Scientific Computing Center, a DOE Office of Science User Facility
supported by the Office of Science of the U.S.\ Department of Energy under
Contract No.\ DE-AC02-05CH11231.  This work was supported in part by NSF Grant
PHY-1307372 at UC San Diego, and LDRD funding at Los Alamos National
Laboratory.  We thank the anonymous referee for their useful comments.

\bibliography{master}

\end{document}

%% file: defs.tex
\newlength{\wdo}
\newcommand{\stroke}[1]{{$#1$}%
\settowidth{\wdo}{${#1}$} {\kern-\wdo}%
\partialvartstrokedint}

\makeatletter
\newcommand{\fancysep}{%
  \@afterindentfalse
  {\begin{center}
    \resizebox{0.8\linewidth}{0.4ex}{{%
        \fontsize{20}{24}\usefont{U}{webo}{xl}{n}{4}}}%
  \end{center}}\@afterheading}
\makeatother

{\vskip 1.0em
\framebox[\columnwidth][r]{%
\begin{minipage}[c]{\columnwidth}%
\vspace{-1.0em}%
#1%
\end{minipage}}}
{\vskip 1.0em}

\def\XXint#1#2#3{{\setbox0=\hbox{$#1{#2#3}{\int}$}
     \vcenter{\hbox{$#2#3$}}\kern-.5\wd0}}

\newcommand{\beq}{\begin{equation}}
\newcommand{\eeq}{\end{equation}}
\newcommand{\beqa}{\begin{eqnarray}}
\newcommand{\eeqa}{\end{eqnarray}}

\newcommand{\eps}{\ensuremath{\epsilon}\xspace}

\newcommand{\nut}{\ensuremath{\nu_\tau}}
\newcommand{\bnut}{\ensuremath{\overline{\nu}_\tau}}

\newcommand{\neff}{\ensuremath{N_\textnormal{eff}}\xspace}

\newcommand{\bardens}{\ensuremath{\omega_b}\xspace}
\newcommand{\burst}{{\sc burst}\xspace}

\newcommand{\num}{\ensuremath{\nu_\mu}\xspace}

\newcommand{\bnum}{\ensuremath{\overline{\nu}_\mu}\xspace}

\newcommand{\tcm}{\ensuremath{T_\textnormal{cm}}\xspace}

\newcommand{\nbins}{\ensuremath{N_\textnormal{bins}}\xspace}
\newcommand{\epsmax}{\ensuremath{\epsilon_\textnormal{max}}\xspace}

\newcommand{\feq}{\ensuremath{f^\textnormal{(eq)}}\xspace}
\newcommand{\nuratiotol}{\ensuremath{\varepsilon({\rm net}/{\rm FRS})}\xspace}
\newcommand{\epsvals}{$\epsilon$ values\xspace}
\newcommand{\epsval}{$\epsilon$-value\xspace}
\newcommand{\tcmpl}{\ensuremath{T_\textnormal{cm}/T}\xspace}
\newcommand{\np}{\ensuremath{n/p}\xspace}

\newcommand{\nue}{{\ensuremath{\nu_{e}}}\xspace}

\newcommand{\bnue}{\ensuremath{\overline{\nu}_e}\xspace}

\newcommand{\spl}{\ensuremath{s_\textnormal{pl}}\xspace}

\newcommand{\ben}{\begin{enumerate}}
\newcommand{\een}{\end{enumerate}}

\newcommand{\sisf}{\ensuremath{s_\textnormal{pl}^{(i)}/s_\textnormal{pl}^{(f)}}\xspace}
\newcommand{\deltadr}{\ensuremath{\delta_\textnormal{dr}}\xspace}
\newcommand{\bnu}{\ensuremath{\overline\nu}\xspace}
\newcommand{\heiv}{\ensuremath{\,^4\textnormal{He}}\xspace}

\newcommand{\lstari}{\ensuremath{L_{i}^{\star}}\xspace}
\newcommand{\lstarnu}{\ensuremath{L_{\nu}^{\star}}\xspace}
\newcommand{\threehe}{\ensuremath{{}^3\textnormal{He}}\xspace}
\newcommand{\livii}{\ensuremath{{}^7\textnormal{Li}}\xspace}
\newcommand{\yp}{\ensuremath{Y_P}\xspace}

\newcommand{\bnui}{\ensuremath{\overline{\nu}_i}\xspace}
\newcommand{\nui}{\ensuremath{\nu_i}\xspace}

\newcommand{\deltamnp}{\ensuremath{\delta m_{np}}\xspace}